\documentclass[%
reprint,
superscriptaddress,
frontmatterverbose, 
preprintnumbers,
longbibliography,
 amsmath,amssymb,
 aps,
 pra,
notitlepage,
nofootinbib,
]{revtex4-1}

\usepackage{amsmath}
\usepackage{braket}
\usepackage{tabularx,graphicx}
\usepackage{epstopdf}
\usepackage{makecell}
\usepackage{graphicx}
\usepackage{latexsym}
\usepackage{amssymb}
\usepackage{amsmath}
\usepackage{color, colortbl}
\usepackage{psfrag}
\usepackage{bbm}
\usepackage{bm}
\usepackage{titlesec}
\usepackage{dsfont}
\usepackage{feynmp}
\usepackage{slashed}
\usepackage{multirow}
\usepackage{url}

\newcommand{\beq}{\begin{eqnarray}}
	\newcommand{\eeq}{\end{eqnarray}}

\newcommand{\bsp}{\begin{aligned}}
	\newcommand{\esp}{\end{aligned}}

\usepackage{xcolor}
\definecolor{darkblue}{rgb}{0.,0.,0.4}
\definecolor{darkred}{rgb}{0.5,0.,0.}
\definecolor{BlueViolet}{RGB}{138,43,226}
\definecolor{SkyBlue}{RGB}{30,144,255}
\definecolor{DarkGreen}{RGB}{0,100,0}

\usepackage[colorlinks=true,linkcolor=darkblue,citecolor=blue,urlcolor=darkred]{hyperref}
\usepackage[normalem]{ulem}

\usepackage{amsthm}
\usepackage{amsmath}
\theoremstyle{plain}

\theoremstyle{definition}

\usepackage{mathrsfs}
\usepackage{comment}

\usepackage{pst-all}
\usepackage{leftidx}
\usepackage[off]{auto-pst-pdf} %%%Speeds up compilation but does not compile any new pstricks figures or recompile old ones.
\usepackage{booktabs}
\usepackage{yfonts}
\makeatletter
\usepackage[caption=false]{subfig}

\begin{document}

\title{
Noise-strength-adapted approximate quantum codes inspired by machine learning
}
	
\author{Shuwei Liu}
\thanks{These authors contributed equally to this work.}
\affiliation{Perimeter Institute for Theoretical Physics, Waterloo, Ontario, Canada N2L 2Y5}
\affiliation{Department of Physics and Astronomy, University of Waterloo, Waterloo, Ontario, Canada N2L 3G1}

\author{Shiyu Zhou}
\thanks{These authors contributed equally to this work.}
\affiliation{Perimeter Institute for Theoretical Physics, Waterloo, Ontario, Canada N2L 2Y5}

\author{Zi-Wen Liu}
\affiliation{Yau Mathematical Sciences Center, Tsinghua University, Beijing 100084, China}

\author{Jinmin Yi}
\affiliation{Perimeter Institute for Theoretical Physics, Waterloo, Ontario, Canada N2L 2Y5}
\affiliation{Department of Physics and Astronomy, University of Waterloo, Waterloo, Ontario, Canada N2L 3G1}

%%%%%%%%%%%%%%%%%%%%%%%%%%%%%%%%%%%%%%%%%%%%%%%%%%%%%%%%%%%%%%%%%%%%%%%%%%%
\begin{abstract}

We demonstrate that machine learning provides a powerful tool for discovering new approximate quantum error-correcting (AQEC) codes beyond conventional algebraic frameworks. Building upon direct observations through hybrid quantum-classical learning, we discover two new 4-qubit amplitude damping codes with an innovative noise-strength-adaptive (NSA) feature where the codeword varies with noise strength. They are NSA self-complementary and NSA pair-complementary codes. We show that they can both outperform conventional codes for amplitude damping (AD) noise. The 4-qubit self-complementary NSA code outperforms the standard LNCY AD code in fidelity and Knill-Laflamme condition violation. The pair-complementary code, which has no known non-NSA analog, achieves even better performance with higher-order loss suppression and better fidelity. We further generalize both approaches to families of NSA AD codes for arbitrary system size, as well as an NSA variant of the 0-2-4 binomial code for single-photon loss. Our results demonstrate that adaptation to noise strength can systematically lead to significant improvements in error correction capability, and also showcase how machine learning can help discover new valuable code formalisms that may not emerge from traditional design approaches.
\end{abstract}

\maketitle

%\tableofcontents

%%%%%%%%%%%%%%%%%%%%%%%%%%%%%%%%%%%%%%%%%%%%%%%%%%%%%%%%%%%%%%%%%%%%%%%%%%%
\section{Introduction} \label{sec:intro} 

Large-scale fault-tolerant quantum computers are essential for achieving the theoretically promised advantages over classical machines in solving certain important problems, and quantum error correction (QEC) plays an essential role for enabling that.
While much research has focused on QEC codes that enable exact recovery of logical information, namely perfect error correction~\cite{shor95,gottesman1997,nielsen2000quantum},  approximate quantum error-correcting (AQEC) codes, which allow for a certain degree of recovery inaccuracy, can provide more practical and sometimes necessary alternatives with important advantages. For example, AQEC can outperform exact QEC in terms of code rates~\cite{Leung_1997,CGS05:aqec} and facilitate continuous transversal logical gates in covariant codes~\cite{Hayden_2021,faist20,Woods2020continuousgroupsof,PhysRevLett.126.150503,PhysRevResearch.4.023107,Zhou2021newperspectives,liu2022approximate,liu2023quantumerrorcorrectionmeets,KongLiu22} which would otherwise be prohibited~\cite{PhysRevLett.102.110502}.

Despite the importance of AQEC codes, our understanding of them remains limited. In particular, there lacks a systematic framework for designing useful AQEC codes, and most known codes depend on pre-assumed algebraic structures, such as stabilizer formalism~\cite{gottesman1997} or self-complementary quantum codes~\cite{Smolin_2007,lang2007}. 
In this work, we propose to employ machine learning (ML) to explore AQEC codes beyond conventional algebraic constraints, thereby expanding the range of possibilities. Specifically, we demonstrate how a hybrid quantum-classical learning algorithm leads to the discovery of novel AQEC codes with unique structures and improved performance. This machine learning approach offers several advantages. Its flexible architecture provides a systematic way to search for AQEC codes suitable for any noise model, while also enables the exploration of non-stabilizer codes.
Although there have been previous works~\cite{Cao_2022, qvector} using variational ML to find QEC codes, little efforts have been devoted to exploring AQEC.

In this study, we demonstrate the discovery of new AQEC codes for amplitude damping (AD) noise---a common noise modeling energy losses in quantum systems that describes photon losses in photonic systems~\cite{QIwithCV2007} and spin relaxations in superconducting systems~\cite{ADsuperconducting}. Through our learning protocol, we find two $(\!(4,1)\!)$ codes capable of approximately correcting a single AD error, with the innovative feature that the code is adaptive to arbitrary AD noise strength (note that Ref.~\cite{mao2024optimizedfourqubitquantumerror} also identifies a different 4-qubit AD code with this feature).  We name such codes {\it noise-strength-adapted (NSA)}  codes.
Namely, the codewords vary as the noise strength varies, allowing the possibility of tailoring the codes towards different quantum device architectures to improve performance. It should be noted that no additional information is required in designing these adaptive codes, since the noise strength value is already necessary for decoding AD codes, as demonstrated in~\cite{Leung_1997}.
Leveraging this property, our codes outperform the LNCY $(\!(4,1)\!)$ AD code~\cite{Leung_1997} by exhibiting lower Knill-Laflamme condition violation and higher worst-case fidelity. 
The two new codes discovered from ML are self-complementary (SC) codes and pair-complementary (PC) codes. The former is an NSA version of the LNCY code, while the latter is an entirely new code. Moreover, the PC code achieves even better performance than the SC code, with higher-order loss suppression and increased fidelity. Note that the code found in Ref.~\cite{mao2024optimizedfourqubitquantumerror} through biconvex optimization can also achieve higher fidelity than the LNCY code~\cite{mao2024optimizedfourqubitquantumerror}, together suggesting that the NSA formalism can consistently improve code performances.

Remarkably, both the SC and PC structures can be systematically generalized into families of AD codes under the framework of NSA, each capable of correcting a single AD error. Compact codeword forms of these families for arbitrary $n$ can also be obtained. We explicitly compute the fidelities for both families of codes as functions of $n$ and $\gamma$, and show that the performance advantage of NSA codes grows with the number of qubits. We also observe that the two families present a trade-off between error-correction performance and code rates. Specifically, for a given $n$, PC code has a smaller $k$ but achieves a higher fidelity. We note that the NSA SC codes can be generalized further to qudit systems with arbitrary local dimension $q \ge 2$.
In addition, our NSA framework can be applied to systems consisting of bosons. One of the simplest examples of a bosonic code, the 0-2-4 binomial code, which can correct a single photon loss~\cite{Binomial2016}, can be adapted to an NSA version with improved fidelity. Our results demonstrate that NSA is an advantageous and comprehensive formalism that may significantly benefit the design of a wide variety of codes.

It is important to emphasize that the NSA $(\!(4,1)\!)$ AD codes originate from our observations in the learning procedures, highlighting the potential of machine learning to identify and refine new codes. Furthermore, we successfully developed an analytical understanding of our NSA codes, enabling systematic generalizations to broader classes of codes.
The remainder of this paper is organized as follows: Sec.~\ref{sec:prem} reviews the principles of approximate quantum error correction and defines amplitude damping noise. Sec.~\ref{sec:experiments} demonstrates how our ML experiments lead to discovering NSA codes, followed by Sec.~\ref{sec:result} where we present our $(\!(4, 1)\!)$ NSA AD code findings. In Sec.~\ref{sec:gen}, we generalize these results by deriving families of $(\!(n, k)\!)$ NSA AD codes and an NSA version of the 0-2-4 binomial code. Finally, Sec.~\ref{sec:conclusion} concludes with summaries and future directions.

%%%%%%%%%%%%%%%%%%%%%%%%%%%%%%%%%%%%%%%%%%%%%%%%%%%%%%%%%%%%%%%%%%%%%%%%%%%
\section{Preliminaries} \label{sec:prem} 
Consider a $2^k$-dimensional subspace of an $n$-qubit Hilbert space as a $(\!(n,k)\!)$ code. While a perfect error correcting code requires the Knill-Laflamme (KL) condition~\cite{KL_condition}, this condition can be relaxed for an approximate code~\cite{BO10:AKL,Ng10}:
\begin{equation}
    \langle \bar\psi_\alpha | E_{a}^\dagger E_{b} | \bar\psi_\beta \rangle = C_{a b} \delta_{\alpha\beta} + \epsilon_{\alpha\beta,ab}
    \; ,
    \label{eq:kl_aqec}
\end{equation}
where $\{| \bar \psi_\alpha \rangle\}$ form the basis of the code subspace, and $\{E_{a}\}$ are the Kraus operators for the noise channel. $\epsilon_{\alpha\beta,ab}$ are small parameters quantifying the KL condition violation and depend on both the code states and the error operators. A code can approximately correct weight-$t$ errors if Eq.~(\ref{eq:kl_aqec}) holds for all error operators with weight up to $t$. The weight of an error operator is defined as the number of qubits on which it acts non-trivially. The Eq.~\eqref{eq:kl_aqec} guarantees that there exists an approximate recovery channel $\mathcal R$ after the encoded information is contaminated by the noise operators, in other words $\mathcal R \circ \mathcal N \circ \mathcal E \approx \text{id}_{\text L}$, where $\mathcal N$ and $\mathcal E$ are the noise and encoding channels, respectively, and $ \text{id}_{\text L}$ denotes the identity channel in the logical subspace. Alternative ways to describe the deviations of an AQEC code from an exact one include the subsystem variance~\cite{Yi_2024} and the near-optimal fidelity~\cite{Zheng_2024}, both of which are related to the violation of the KL condition in Eq.~(\ref{eq:kl_aqec}).

One can also benchmark the performance of an AQEC code by the worst-case fidelity, which we will simply refer to as fidelity later on,
\begin{equation}
    \mathcal{F}=\max_{\mathcal{R}}\left\{ \min_{|\bar\psi\rangle} \,
    \langle\bar\psi| \,
    (\mathcal{R}\circ\mathcal{N})(|\bar\psi\rangle\langle\bar\psi|)
    \,
    |\bar\psi\rangle \right\}
    \; ,
\end{equation}
where minimization is taken over the code states $|\bar\psi\rangle$ and maximization is taken over all possible recovery channels $\mathcal{R}$. While in general there is no restriction on the recovery channel, we restrict ourselves to $\mathcal{R}$ such that after the recovery there will be no residual logical error. Note that the scaling of $1-\mathcal{F}$ with respect to the noise strength $\gamma$ is lower-bounded by the KL condition violation’s scaling (see the proof in supplementary materials).

In this work, we aim to use machine learning to find AQEC codes suitable for amplitude damping noise, which in a two-level system corresponds to the spontaneous decay from the excited state to the ground state. The noise channel $\mathcal{N}^{A}(\rho)=A^0\rho A^{0 \dagger}+A^1\rho A^{1 \dagger}$ is given by the Kraus operators $ A^0 = |0\rangle\langle0|+\sqrt{1-\gamma}|1\rangle\langle1|$ and $A^1=\sqrt{\gamma}|1\rangle\langle0|$, where the decay rate $\gamma$ characterizes the noise strength. $A^1$ corresponds to a `jump' from the state $| 1 \rangle$ to $| 0 \rangle$, while $A^0$ reduces the amplitude of $| 1 \rangle$ by the amount $\sqrt{1 - \gamma}$. We consider the following independent amplitude damping noise model, $\mathcal{N}= \otimes_{i=1}^n \mathcal{N}_i^A$, where $\mathcal{N}_i^A$ is the amplitude damping channel acting on the $i$-th qubit with Kraus operators $A^0$ and $A^1$. Note that each qubit is affected independently by its own noise operator.

%%%%%%%%%%%%%%%%%%%%%%%%%%%%%%%%%%%%%%%%%%%%%%%%%%%%%%%%%%%%%%%%%%%%%%%%%%%

\section{Machine learning experiments} \label{sec:experiments}

In the following, we demonstrate how our machine learning experiments played a key role in discovering NSA codes for amplitude damping noise. While our NSA formulation is motivated by experimental results, its analytical construction remains rigorous and independent of the learning process.

To search for AQEC codes, we employ a variational quantum learning approach, where the bases of the code space are generated by a parametrized quantum circuit. This circuit consists of multiple layers of one- and two-qubit rotation gates with trainable variables. For an AQEC code designed to approximately correct $t$ errors, the KL condition violation is required to scale as $O(\gamma^{t+1})$~\cite{Kong2015}, where $\gamma$ denotes the noise strength. To enforce this scaling, we define our loss function in our variational quantum learning as the sum of the $L^1$-norms of the KL condition violations~\cite{Cao_2022}:
\begin{align}
    \mathcal L = 
    \sum_{\mu} 
    &\left[
    \sum_{1 \leqslant \alpha < \beta \leqslant 2^k} |\langle \bar\psi_\alpha | E_{\mu} | \bar\psi_\beta \rangle |\right]
    \nonumber
    \\
    &+ \sum_{\mu} \left[ \sum_{\alpha = 1}^{2^k} \frac{1}{2} \left| \langle \bar\psi_\alpha | E_{\mu} | \bar\psi_\alpha \rangle - \overline{\langle E_{\mu} \rangle} \right| 
    \right]
     \; , 
    \label{eq:loss1}
\end{align}
where $E_{\mu} = E_a^\dagger E_b$ and each Kraus operator $E_a$ is drawn from the error set $\{E_a\}$. 
Our experiments focus on a 4-qubit system designed to approximately correct a single amplitude damping error $(t=1)$. The independent error set consists of operators $E_a = \otimes_{i=1}^n A_i^{l_i}$, where at most one qubit experiences a jump error $A^1$, while the remaining qubits are affected by $A^0$.

\begin{figure}[t]
    \centering
      \includegraphics[width=\linewidth]{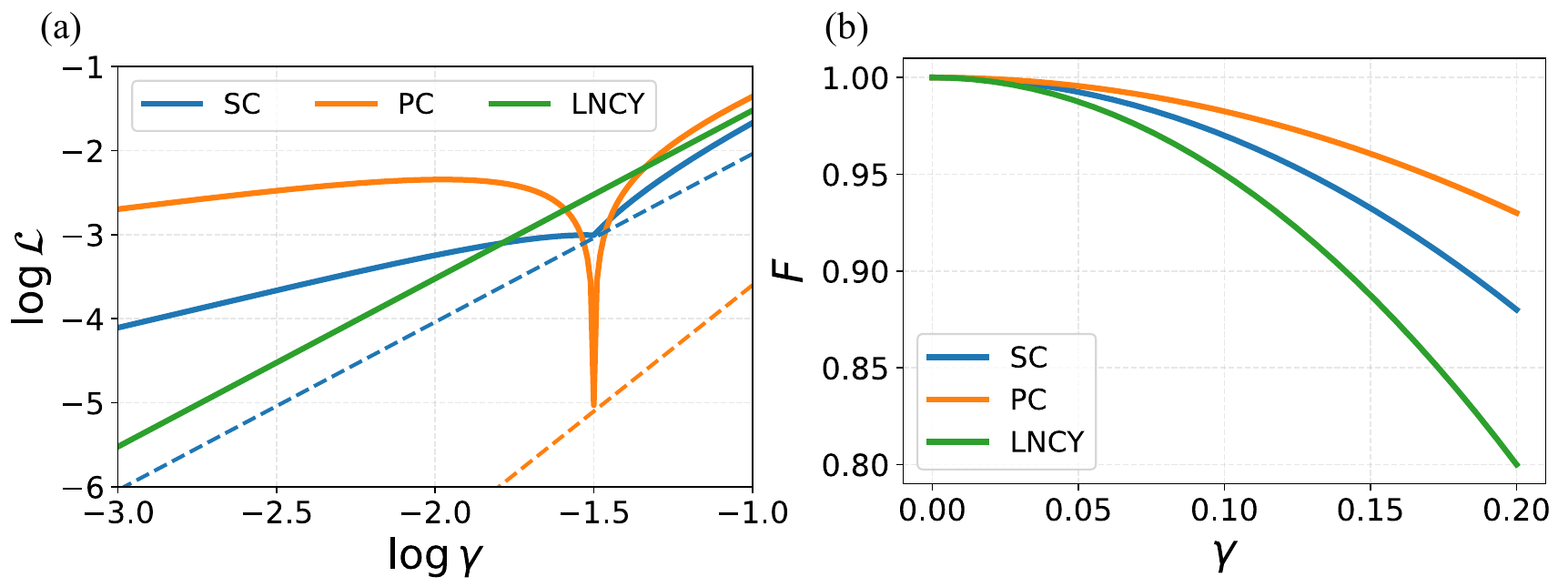}
    \caption{
    (a) Loss $\mathcal L$ as a function of $\gamma$ is shown. The solid green line shows the smooth $\gamma^2$ dependence of LNCY code. The solid blue and orange lines show the losses of two codes optimized for $\gamma=10^{-1.5}$. The blue and orange dashed lines, instead, show the losses for NSA codes with adaptive code states.
    (b) shows the fidelity plotted as a function of $\gamma$ for $(\!(4,1)\!)$ NSA SC, $(\!(4,1)\!)$ NSA PC, and LNCY codes. 
    }
    \label{fig:combined}
\end{figure}

We begin with applying the
variational quantum learning protocol to search for the best code states at a fixed noise strength $\gamma_0$ by minimizing %the loss function defined in 
Eq.~(\ref{eq:loss1}). The details of the learning procedure are provided in the supplementary material. After obtaining an optimized code at $\gamma_0$, we evaluate its performance across a range of noise strengths $\gamma$, as shown in Fig.~\ref{fig:combined}(a). In general, for an AQEC code correcting a single AD noise, the loss function is expected to vary smoothly with $\gamma$ and typically scales as $O(\gamma^{2})$, ensuring that the code remains effective for a general strength $\gamma$. This expected behavior is exhibited in the $(\!(4,1)\!)$ LNCY code in~\cite{Leung_1997}, as shown in the green line in Fig.~\ref{fig:combined}(a). 

However, the codes found through our variational learning optimized for a specific noise strength $\gamma_0$ exhibit a strikingly different pattern. Not only do they achieve lower losses around $\gamma_0$ compared to the LNCY code, but their loss functions also reveal a distinct {\it non-smooth point} precisely at $\gamma = \gamma_0 = 10^{-1.5}$, as shown in the orange and blue solid lines in Fig.~\ref{fig:combined}(a).  
This observation leads to significant implications: i)
the codes learned by our variational protocol must be fundamentally different from the LNCY code; ii) AQEC codes optimized for a specific noise strength 
can not be realized as a single fixed-parameter code across varying noise regimes. Instead, the optimal code structure itself should vary with $\gamma$. This motivates us to introduce a new code construction---noise-strength adapted (NSA) codes---in which the code states are designed to continuously adjust with the noise strength $\gamma$. 

As shown in Fig.~\ref{fig:combined}(a), the machine learning results reveal the presence of two saddle points, indicated in the blue and orange lines. Both saddle points correspond to code structures with reduced loss around $ \gamma_0$, and they belong to two distinct families of codes. With the new NSA construction, we demonstrate that the loss function for codes with varying code states smoothly changes with $\gamma$,
as indicated in the dashed lines in Fig.~\ref{fig:combined}(a). We demonstrate in the next section that these two families—termed self-complementary and pair-complementary codes—differ in their structural properties (with the naming convention becoming clear from the analytical formulation), and offer better performance over a broad range of noise strength, compared to the conventional non-NSA approximated code.

%%%%%%%%%%%%%%%%%%%%%%%%%%%%%%%%%%%%%%%%%%%%%%%%%%%%%%%%%%%%%%%%%%%%%%%%%%%
\section{Basic analytical results}\label{sec:result}

Our machine learning experiment on 
$4$-qubit codes reveals a distinct structured pattern in the code states, which can be captured analytically by a closed-form expression that adapts to the noise strength $\gamma$. 

We begin by presenting the 4-qubit NSA self-complementary code:
\begin{align}
    | \bar 0 \rangle
		&=
		\sqrt{\frac{1}{1+(1-\gamma)^{-4}}}(|0000\rangle+(1-\gamma)^{-2}|1111\rangle)
	\; ;
	\nonumber 
	\\
    | \bar 1 \rangle
		&=
		\sqrt{\frac{1}{2}}\left(|0011\rangle+|1100\rangle\right)
	\; .
\label{eq:nsa_4qubit}
\end{align}
This code differs from LNCY code in that the coefficients in $| \bar0 \rangle$ depend on $\gamma$, whereas the coefficients in both $| \bar0 \rangle$ and $| \bar1 \rangle$ in LNCY code are fixed at $1/\sqrt{2}$. In the limit of $\gamma\rightarrow 0$, the NSA code reduces to its non-NSA counterpart. 

The loss $\mathcal L$ for the NSA code in Eq.~(\ref{eq:nsa_4qubit}) is $\gamma^2+O(\gamma^3)$, which is smaller compared to the LNCY code, $3\gamma^2+O(\gamma^3)$. Furthermore, without allowing residual logical errors, the fidelity for the NSA code is $1 - 3\gamma^2 + \mathcal{O}(\gamma^3)$, cf.~$ 1 - 5\gamma^2 + \mathcal{O}(\gamma^3)$ for the LNCY code, plotted as the blue and green lines in Fig.~\ref{fig:combined}(b) respectively. The difference in the fidelities lies in that the recovery channel distorts the encoded states in LNCY non-NSA code, but such distortion can be absent in our NSA code, see supplementary materials for details.  Notably, even in the non-NSA code, where the codewords are independent of $\gamma$, knowledge of the noise strength remains crucial for decoding~\cite{Leung_1997}. 

Next, we introduce another 4-qubit NSA code, termed the pair-complementary (PC) code, which differs from the self-complementary code by having code states consisting of four basis states:
\begin{align}
    | \bar0 \rangle &= \left( 
        (1-\gamma)^{-1} | 0011 \rangle -
        (1-\gamma)^{-3/2} | 1110 \rangle \right.
        \nonumber \\
        & \left.- (1-\gamma)^{-3/2} | 1101 \rangle +
        | 0000 \rangle
    \right) / N_0 \; ; \nonumber \\
     | \bar1 \rangle &= \left( 
        (1-\gamma)^{-1} | 1100 \rangle +
        (1-\gamma)^{-1/2} | 0001 \rangle \right.
        \nonumber \\
        &\left.+ (1-\gamma)^{-1/2} | 0010 \rangle +
        (1-\gamma)^{-2} | 1111 \rangle 
    \right) / N_1
    \; ,
    \label{eq:dc_4}
\end{align}
where $N_0$ and $N_1$ are normalizations. Notice that the basis states involved in the two logical states are pairwise complementary to each other. The loss for the pair-complementary code is $\gamma^3/4 + O(\gamma^4)$, and achieves higher-order loss suppression compared to the self-complementary codes, as also illustrated by the dashed orange lines in Fig.~\ref{fig:combined}(a). 
Additionally, the fidelity for the pair-complementary code, given by $1-7\gamma^2/4 + O(\gamma^3)$ and marked as orange line in Fig.~\ref{fig:combined}(b), is larger than the NSA 4-qubit self-complementary code. Notably, this new code has no corresponding non-NSA versions that have been reported in the literature. 

Both codes derived from ML experiments can be generalized to arbitrary system size. In the following section, we introduce these two families of NSA codes and demonstrate their enhanced performance in terms of fidelity.

%%%%%%%%%%%%%%%%%%%%%%%%%%%%%%%%%%%%%%%%%%%%%%%%%%%%%%%%%%%%%%%%%%%%%%%%%%%
\section{Generalizations} \label{sec:gen}
As observed in the 4-qubit NSA codes, the optimized code structure introduces asymmetry into the traditional SC codes which admit a basis of codewords of the form $| \bar \psi_u \rangle = (| u \rangle + | \tilde u \rangle)/\sqrt{2}$, where for each qubit \( i \), \( \tilde u_i = 1 - u_i \). The key insight from the machine learned SC codes is that allowing different coefficients for \( | u \rangle \) and its complement, adjusted based on the noise strength, enhances code performance. 

Expanding on this concept, we extend the formalism to general $(\!(n,k)\!)$ NSA SC codes\footnote{We assume the largest possible $k$ for a given $n$ qubit systems The fidelity calculations conducted later are also under the same assumption.}, building upon an existing family of $(\!(n,k)\!)$ SC codes designed to correct $t=1$ amplitude damping noise~\cite{lang2007}, and arrive at the following structure for the codeword: 
\begin{equation}
    | \bar \psi_u \rangle 
        = (1 - \gamma)^{ - \|u\|/2}|u \rangle 
        + (1 - \gamma)^{ - \|\tilde u\|/2}|\bar u \rangle 
        / N_u
    \; ,
    \label{eq:nsa_qubit}
\end{equation}
where $N_u$ is the normalization, and for a given bit string $u$, $\|u\|$ denotes the Hamming weight. Notice that each basis coefficient is a function of the noise strength $\gamma$ and the basis itself. We then compute the fidelity for the NSA SC codes and find 
\begin{align}
    \mathcal F_{\text{SC}}
        = 1 - (n^2-n) \, \gamma^2 / 4 + \mathcal{O}(\gamma^3)
    \; .
    \label{eq:f_qudit}
\end{align}
We compare it with the non-NSA codes and find that up to $O(\gamma^2)$ NSA codes achieve a higher fidelity by the amount given by $(2n\lfloor n/2 \rfloor - 2 \lfloor n/2 \rfloor^2)\gamma^2 / 4$. Note that the performance enhancement of NSA codes grows as the system size increases. In addition, the NSA SC code can be readily extended to qudit systems with a local dimension $q \ge 2$, and for further details, including the fidelity calculations, we refer readers to the supplementary material.

We now turn to discuss the generalization of pair-complementary NSA codes. Building on our finding of $(\!(4,1)\!)$ NSA PC code through machine learning, we show that there exists a new family of $(\!(n,k)\!)$ code to correct a single amplitude damping noise. 
The codeword takes the following form,
\begin{align}
    |\bar\psi_u\rangle
        &\sim |u11\rangle+|u00\rangle-|\tilde{u}10\rangle-|\tilde{u}01\rangle \nonumber \; ;\\
    |\bar{\psi}'_{u}\rangle
        &\sim |\tilde{u}00\rangle+
        |\tilde{u}11\rangle+
        |u01\rangle+
        |u10\rangle
        \; ,
\end{align}
where all coefficients and normalizations are omitted to emphasize the code structure more clearly. The coefficients, which follow the same form as in Eq.~\eqref{eq:nsa_qubit}, depend on both $\gamma$ and the Hamming weight of the basis.  Notice that the codewords of this new family is constructed by extending the basis $u$ in the SC codeword with two additional bits. Specifically, the $|\bar{\psi}_{u}\rangle$ and $|\bar{\psi}'_{u}\rangle$ pair contains superpositions of all combinations $\{0,1\}^2$ appending after the SC codeword $u$ and its complementary $\tilde u$. Hence, we claim that for any given $(\!(n,k)\!)$ SC code, there exists a corresponding $(\!(n+2,k+1)\!)$ PC code that can correct one amplitude damping error (see supplementary for proof details).

We compute the fidelity of NSA PC codes for general $n$, and obtain
\begin{align}
    \mathcal F_{\text{PC}} = 1-(n^2-3n+3)\gamma^2/4+O(\gamma^3) \;.
\end{align}
We observe that for any given $n$, pair-complementary NSA codes achieve higher fidelity than the self-complementary ones, at the cost of fewer logical qubits. Specifically, we establish the following inequality, $\mathcal{F}_\text{SC}[(\!(n,k)\!)] > \mathcal{F}_\text{PC}[(\!(n+2,k+1)\!)]>\mathcal{F}_\text{SC}[(\!(n+2,k+1)\!)]$.

The formalism of NSA can be further extended to the bosonic codes~\cite{Binomial2016}. For the simplest 0-2-4 binomial code, with the code states: $|\bar{0'}\rangle=\frac{1}{\sqrt{2}}\left(|0\rangle+|4\rangle\right),~|\bar{1'}\rangle=|2\rangle$, it can be generalized to the NSA version, with $|\bar 0'\rangle$ becoming
\begin{equation}
     |\bar{0}\rangle=\left(|0\rangle+(1-\gamma)^{-2}|4\rangle\right)
    /\sqrt{1+(1-\gamma)^{-4}}
    \; ,
\end{equation}
while $|\bar{1}\rangle=|2\rangle$ remains. Note that this code can be reduced from the $(\!(4,1)\!)$ qubit code by summing up the bit strings in each basis. Hence, the behavior of the fidelity is also the same, which is $1 - 3\gamma^2 + \mathcal{O}(\gamma^3)$ for the NSA code, as compared to $ 1 - 5\gamma^2 + \mathcal{O}(\gamma^3)$ in the non-NSA code.

%%%%%%%%%%%%%%%%%%%%%%%%%%%%%%%%%%%%%%%%%%%%%%%%%%%%%%%%%%%%%%%%%%%%%%%%%%%

\section{Discussions} \label{sec:conclusion}

Building on observations from machine learning methods, we formalized a novel framework for AQEC codes featuring noise strength adaptation, which enables advantages over conventional codes. Specifically, by employing a hybrid quantum-classical learning algorithm, we identified two $(\!(4, 1)\!)$ noise-strength-adapted AQEC codes tailored for amplitude damping noise, with performance surpassing its previously known non-NSA counterpart. The NSA framework generalizes to two families of codes for arbitrary system size, showcasing its scalability. Moreover, it extends to AQEC for single-photon loss through the NSA version of the 0-2-4 binomial code.

Our NSA-based codes optimize code performance by directly incorporating the actual noise strength into the design process, opening a new avenue for code design. Our findings show that NSA codes consistently outperform their non-NSA counterparts, highlighting the advantages of adapting codes to specific noise environments. This suggests that quantum error correction and fault tolerance procedures may significantly benefit from noise-strength adaptation. Future investigations could explore how the NSA principle may extend to other noise models or hybridize with existing QEC paradigms to enhance performance under real-world conditions. 
Furthermore, loss functions based on different measures~\cite{Zheng_2024, mao2024optimizedfourqubitquantumerror} can be employed, which may lead to other structures of NSA codes and inspire new AQEC studies.
 
In this study, the enhanced performance of our NSA codes relies on the constraint that no residual logical errors remain after correction. Nevertheless, we anticipate that our NSA framework can also be effectively applied to scenarios where this constraint is not strictly necessary. Additionally, we generalized the NSA self-complementary code to qudit systems and expect similar generalizations to be feasible for pair-complementary codes. 

Beyond NSA codes, our ML-based methodology provides a systematic approach for exploring quantum error correction codes. In particular, for codes with well-defined algebraic structures—such as stabilizer codes— machine learning methods, including reinforcement learning, could be effective in identifying new codes~\cite{olle_simultaneous_2024,su2023discoveryoptimalquantumerror,Liu2025RLQEC}. By applying these techniques to structured families of quantum codes, future research may uncover novel error correction strategies that would be difficult to find through conventional analytical approaches.

\begin{acknowledgments}
We thank Beni Yoshida for helpful discussions {and Victor Albert for bringing the recent relevant work of Mao et al.~\cite{mao2024optimizedfourqubitquantumerror} to our attention}.
Research at Perimeter Institute is supported in part by the Government of Canada through the Department of Innovation, Science and Industry Canada and by the Province of Ontario through the Ministry of Colleges and Universities. Z.-W.L.\ is supported in part by a startup funding from YMSC, Tsinghua University, and NSFC under Grant No.~12475023.
\end{acknowledgments}

\bibliography{ref.bib}

\end{document}

\title{Supplementary Material: \\Noise-strength-adapted approximate quantum codes inspired by machine learning}
\author{Shuwei Liu}
\thanks{These authors contributed equally to this work.}
\affiliation{Perimeter Institute for Theoretical Physics, Waterloo, Ontario, Canada N2L 2Y5}
\affiliation{Department of Physics and Astronomy, University of Waterloo, Waterloo, Ontario, Canada N2L 3G1}
\author{Shiyu Zhou}
\thanks{These authors contributed equally to this work.}
\affiliation{Perimeter Institute for Theoretical Physics, Waterloo, Ontario, Canada N2L 2Y5}
\author{Zi-Wen Liu}
\affiliation{Yau Mathematical Sciences Center, Tsinghua University, Beijing 100084, China}
\author{Jinmin Yi}

\affiliation{Perimeter Institute for Theoretical Physics, Waterloo, Ontario, Canada N2L 2Y5}
\affiliation{Department of Physics and Astronomy, University of Waterloo, Waterloo, Ontario, Canada N2L 3G1}
\maketitle
%%%%%%%%%%%%%%%%%%%%%%%%%%%%%%%%%%%%%%%%%%%%%%%%%%%%%%%%%%%
\onecolumngrid
%\appendix

%%%%%%%%%%%%%%%%%%%%%%%%%%%%%%%%%%%%%%%%%%%%%%%%%%%%%%%%%%%
\section{Variational quantum learning} 
\label{sec:vqc}

In this section, we provide details of our variational quantum learning approach and explain how the properties of the logical states found through machine learning (ML) guided us in developing our noise-strength adapted (NSA) code.

\begin{figure}[htbp]
    \centering
    \includegraphics[width=0.5\linewidth]{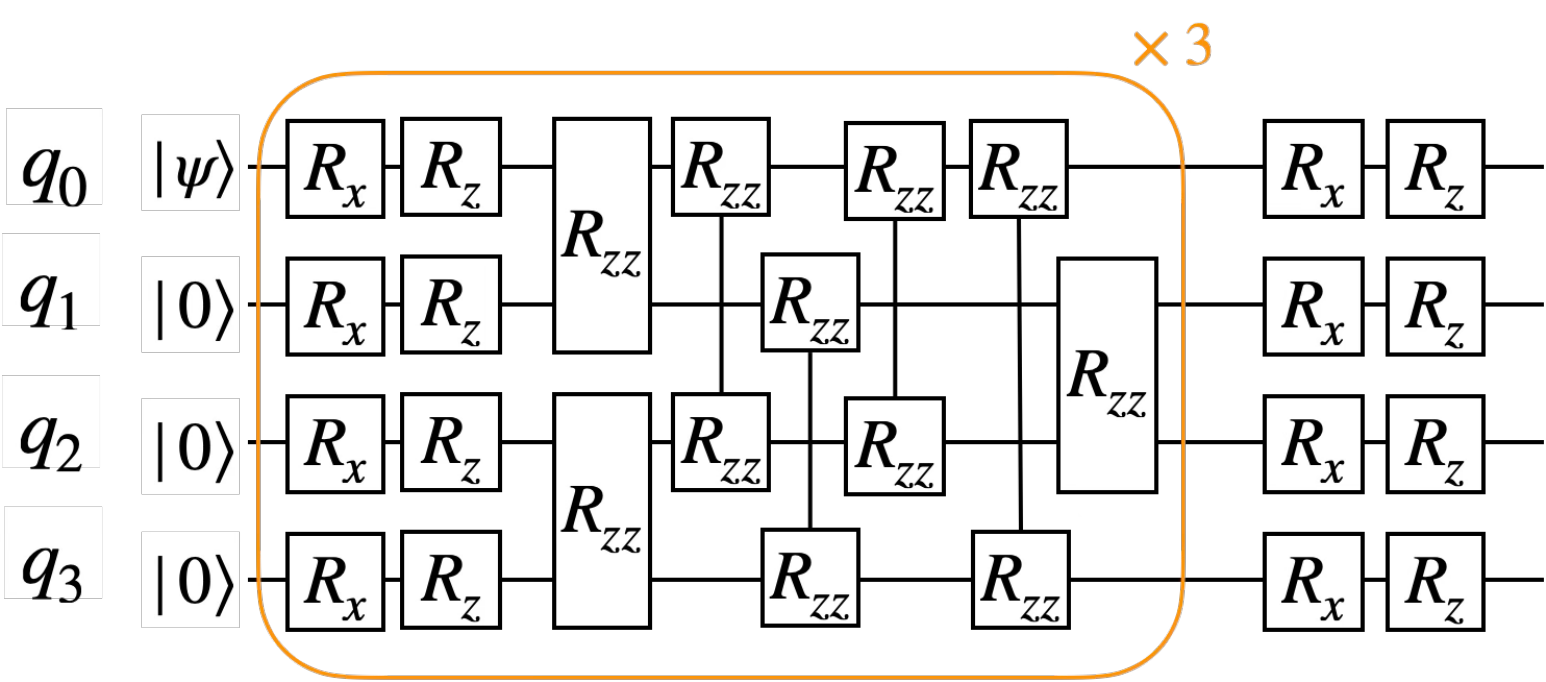}
    \caption{Encoding circuit with trainable parameters for searching \( (\!(4,1)\!) \) code.}
    \label{fig:vqc}
\end{figure}

To identify the code subspace, we employ a parameterized quantum circuit introduced in~\cite{Cao_2022}. This circuit comprises multiple layers of one- and two-qubit rotation gates with trainable parameters, as shown in Fig.~\ref{fig:vqc}. For an $(\!(n,k)\!)$ code, the first $k$ qubits are initialized as the logical state we want to encode, and the rest of the qubits are initialized as the $|0\rangle$ states. For instance, if $k=1$, the first qubit are initialized as $|0\rangle$ and $|1\rangle$ for the logical states $|\bar{0}\rangle$ and $|\bar{1}\rangle$ respectively, while all other states are initialized as the $|0\rangle$ states. 
After initialization, we apply multiple layers of rotation gates, $\{R_x, R_z, R_{zz}\}$, to encode the states. Here, $R_x$ and $R_z$ are single-qubit rotation gates about the $x$- and $z$-axes, respectively, while $R_{zz}$ is a two-qubit entangling gate defined as $e^{-i \theta \sigma^z_i \sigma^z_j / 2}$. This gate set is universal, as any unitary operation can be decomposed into a combination of these three gates.

Specifically, our circuit consists of three layers of rotation gates. Within each layer, we apply $R_x$ and $R_z$ rotations to each physical qubit and $R_{zz}$ rotations to all possible qubit pairs. Each rotation gate has a trainable rotation angle, resulting in a total of $(3n^2 + 13n)/2$ parameters. In our experiments, we focus on searching for the $(\!(4,1)\!)$ code using a 50-parameter parameterized quantum circuit.

As discussed in the main text, an approximate quantum error correcting (AQEC) code designed to approximately correct \( t = 1 \) errors requires the violation of the KL condition to scale as \( O(\gamma^2) \)~\cite{Kong2015}. To enforce this scaling, we define the \( \mathcal L_1 \) loss function in our variational quantum learning framework as the sum of the \( L^1 \)-norms of the KL condition violations. To enhance training stability, we introduce an additional \( \mathcal L_2 \) loss function, also defined as the sum of the \( L^2 \)-norms of the KL condition violations. The training procedure consists of two stages: we first perform pre-training using the \( \mathcal L_2 \) loss function, followed by fine-tuning with the \( \mathcal L_1 \) loss function~\cite{Cao_2022}. The loss functions are defined as follows:
\begin{align}
    \mathcal L_1 = 
    \sum_{\mu} 
    &\left[
    \sum_{1 \leqslant \alpha < \beta \leqslant 2^k} |\langle \bar\psi_\alpha | E_{\mu} | \bar\psi_\beta \rangle |\right]
    + \sum_{\mu} \left[ \sum_{\alpha = 1}^{2^k} \frac{1}{2} \left| \langle \bar\psi_\alpha | E_{\mu} | \bar\psi_\alpha \rangle - \overline{\langle E_{\mu} \rangle} \right| 
    \right]
     \; ,  
     \label{eqn:loss1}\\
   \mathcal L_2 = 
    \sum_{\mu} 
    &\left[
    \sum_{1 \leqslant \alpha < \beta \leqslant 2^k} |\langle \bar\psi_\alpha | E_{\mu} | \bar\psi_\beta \rangle |^2\right]
    + \sum_{\mu} \left[ \sum_{\alpha = 1}^{2^k} \frac{1}{4} \left| \langle \bar\psi_\alpha | E_{\mu} | \bar\psi_\alpha \rangle - \overline{\langle E_{\mu} \rangle} \right| ^2
    \right]
     \; .
\end{align}
Here $\{\ket{\bar\psi_\alpha}\}$ are encoded states after parameterized quantum circuit in Fig.~\ref{fig:vqc} , $E_{\mu} = E_a^\dagger E_b$, and each Kraus operator is drawn from the independent error set $E_a= \otimes^n_{i=1} A_i^{l_i}$ , where at most one qubit experiences a jump error ($A^1$), while the remaining qubits
are affected by $A^0$.

In our experiments, we used the BFGS method from the SciPy optimization package (scipy.optimize) as our optimizer. Within 20,000 steps, our minimization process successfully identified optimized encoding circuits for the code subspace. As discussed and illustrated in Fig. 1 of the main text, the codes discovered through our variational quantum learning are optimized for a specific noise strength $\gamma_0$. However, while these codes outperform known codes ~\cite{Leung_1997} within certain regimes around this specific noise strength $\gamma_0$, they perform worse in other regimes. This suggests that AQEC codes optimized for a specific noise strength cannot performs optimally across different noise regimes. Instead,
the optimal code structure itself should vary with $\gamma$.

\begin{figure}[htbp]
    \centering
    \includegraphics[width=0.9\linewidth]{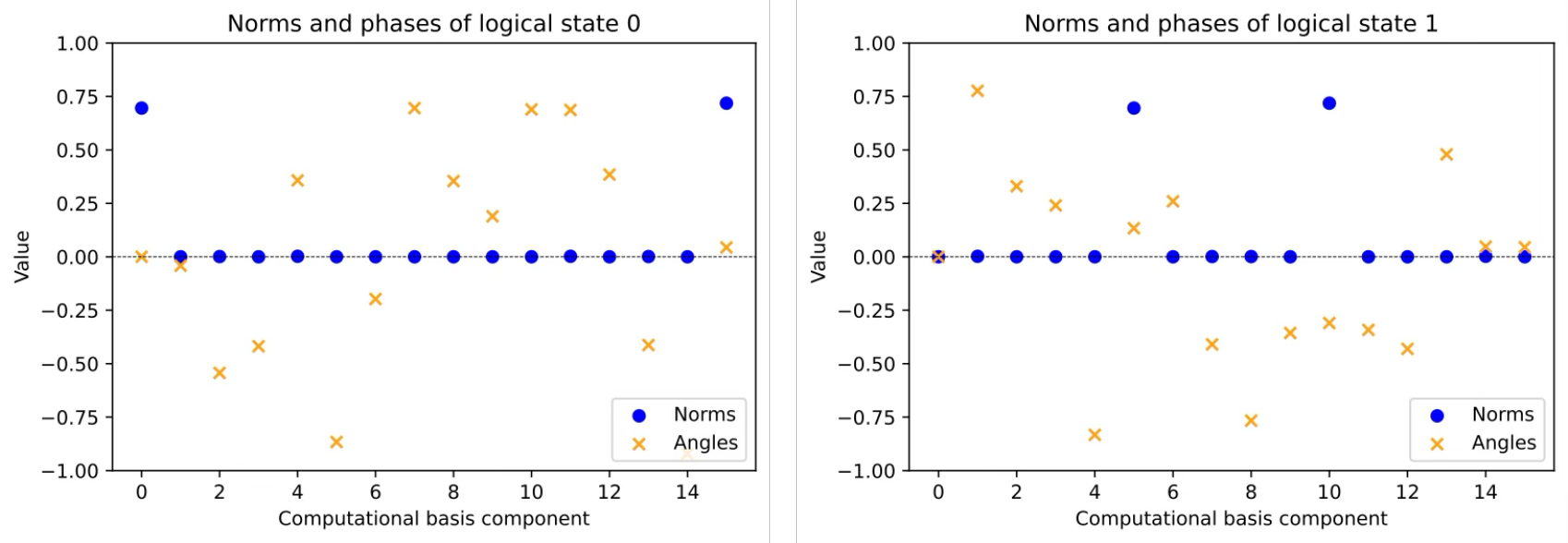}
    \caption{Codeword obtained through variational quantum learning, corresponding to the self-complementary NSA code.}
    \label{fig:ML_codeword_SC}
\end{figure}

To investigate how codewords vary with noise strength $\gamma$, we analyzed the norms and angles of codewords obtained through our variational quantum learning method at $\gamma=\gamma_0 = 10^{-1.5}$. The codewords corresponding to the blue line in Fig. 1 of the main text are illustrated in Fig.~\ref{fig:ML_codeword_SC} above. We observe that only two computational basis components have a considerable norm, and their phases appear to be quite random. Upon closer examination of the norms, we find that their differences are proportional to certain factors of \( \gamma_0 \).  
Therefore, we propose the following ansatz for the codewords:
\begin{align}
    \ket{\bar 0 } &= A(\gamma)\ket{0000}+\sqrt{1-A(\gamma)^2}\ket{1111} \; ;\nonumber \\
        \ket{\bar 1 } &= B(\gamma)\ket{0011}+\sqrt{1-B(\gamma)^2}\ket{1100}\;.
    \label{eq:cw_ansatz}
\end{align}
We further compute the \( L_1 \) loss of these codewords as a function of the coefficients \( A(\gamma) \) and \( B(\gamma) \). Our results show that the choice \( A(\gamma) = 1/{\sqrt{1+(1-\gamma)^{-4}}} \) and \( B(\gamma) = 1/{\sqrt{2}} \) yields a minimum of the loss function in Eq.~(\ref{eqn:loss1}), leading to the self-complementary NSA code presented in Eq.~(4) of the main text. 

\begin{figure}[htbp]
    \centering
    \includegraphics[width=0.9\linewidth]{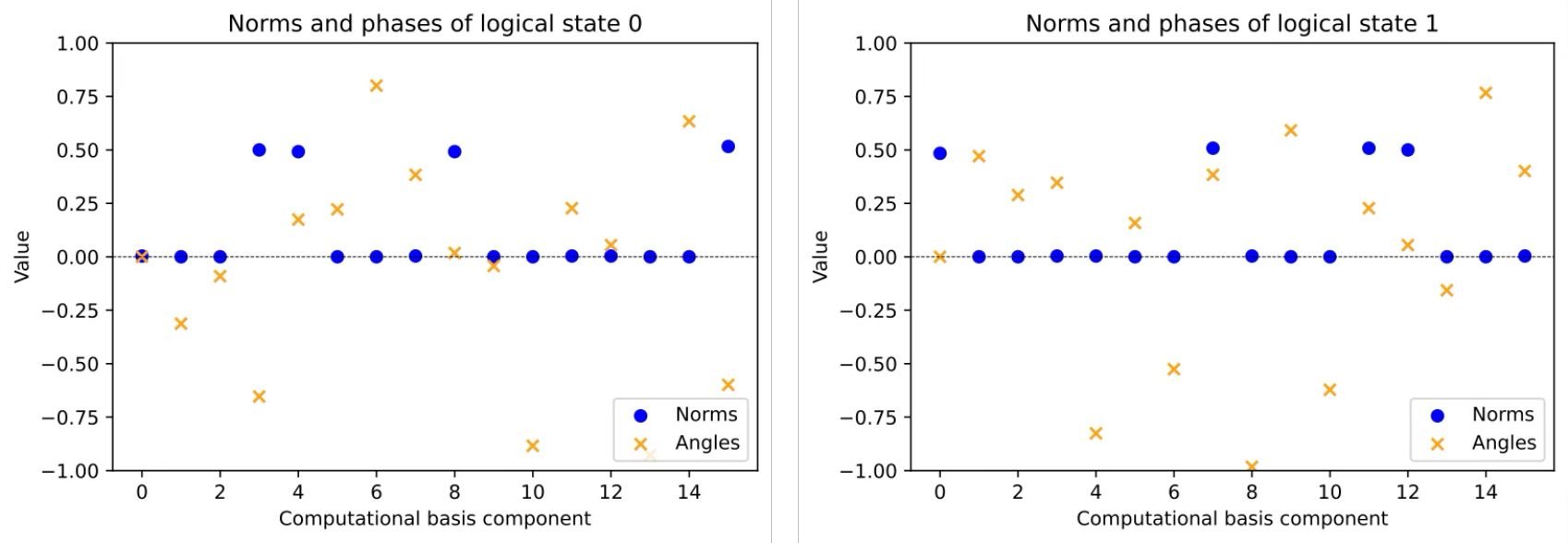}
    \caption{Codeword obtained through variational quantum learning, corresponding to the pair-complementary NSA code.}
    \label{fig:ML_codeword_CC}
\end{figure}

We perform the same analysis for a second code learned from ML, corresponding to the orange line in Fig.~1 in the main text. The optimized codewords from ML are shown in Fig.~\ref{fig:ML_codeword_CC}.
Similarly, we find that when parameterizing the codeword basis as a function of $\gamma$ as in Eq.~\ref{eq:cw_ansatz}, we obtain a code of the form given by Eq.~(5) in the main text, which yields a  minimum of the loss function.

%%%%%%%%%%%%%%%%%%%%%%%%%%%%%%%%%%%%%%%%%%%%%%%%%%%%%%%%%%%
\section{Worst-case fidelity and loss function for AQEC codes}
In this section, we provide detailed information on the relationship between the worst-case fidelity and the loss function that we discuss in the main text. Specifically, we will define three types of measures that describe the deviation from the exact QEC: the completely bounded fidelity $\mathcal{F}_c$, the worst-case fidelity $\mathcal{F}$, and the loss function $\mathcal{L}$. We will focus on the scaling of these measures in terms of the noise strength $\gamma$, and we prove that the scaling of the loss function $\mathcal{L}$ provides a lower bound for the scaling of $1-\mathcal{F}$ and $1-\mathcal{F}_c$, i.e. if $\mathcal{L}=O(\gamma^{t+1})$ where $\gamma$ is the noise strength, then $1-\mathcal{F}_{c}=O(\gamma^{t+1})$ and $1-\mathcal{F}=O(\gamma^{t+1})$.

We now define these two fidelity measures individually:
\begin{defn}[Completely bounded fidelity]
For encoding map $\mathcal{E}$ and noise channel $\mathcal{N}$ acting on the physical system, the \emph{Completely bounded fidelity} is defined as 
    \begin{equation}
 \mathcal{F}_c\coloneqq\max_{\mathcal{R}}\min_{\rho}f((\mathcal{\mathcal{R} \circ \mathcal{N} \circ \mathcal{E}} \otimes \id)(\rho),( \id_L \otimes \id)(\rho))
 \; ,
\end{equation}
where $f$ is the Uhlmann fidelity $f(\rho, \sigma)\coloneqq\operatorname{Tr} \sqrt{\sqrt{\rho} \sigma \sqrt{\rho}}$, and $\id_L$ denotes the logical identity channel. The optimization involves $\mathcal{R}$, which runs over recovery channels, and $\rho$, which runs over input states on any extended system.
\end{defn}

\begin{defn}[Worst-case fidelity]
For encoding map $\mathcal{E}$ and noise channel $\mathcal{N}$ acting on the physical system, the \emph{worst-case fidelity} is defined as 
    \begin{equation}
 \mathcal{F}\coloneqq\max_{\mathcal{R}}\min_{\rho}f(\mathcal{\mathcal{R} \circ \mathcal{N} \circ \mathcal{E}} (\rho),\rho)
 \; ,
\end{equation}
where $f$ is the Uhlmann fidelity $f(\rho, \sigma)\coloneqq\operatorname{Tr} \sqrt{\sqrt{\rho} \sigma \sqrt{\rho}}$, and $\id_L$ denotes the logical identity channel. The optimization involves $\mathcal{R}$, which runs over recovery channels, and $\rho$, which runs over input logical states.
\end{defn}

Note that $\mathcal{F}_c\leq\mathcal{F}$, so we just need to prove that if $\mathcal{L}=O(\gamma^{t+1})$ where $\gamma$ is the noise strength, then $1-\mathcal{F}_c=O(\gamma^{t+1})$. Our proof is based on the following lemma:

\begin{lem}[Proposition 2 in~\cite{Cao_2022}]
Consider an n-qubit noise channel $\mathcal{N}(\rho)=\sum_a E_a \rho E_a^{\dagger}$, and a quantum error correcting code spanned by code words $\{\ket{\psi_1},...,\ket{\psi_K}\}$. Denote the cost function of the basis states as $\mathcal{L}$. Then if the code is $\varepsilon$-correctable under $\mathcal{N}$, i.e. $\sqrt{1-\mathcal{F}_c}=\varepsilon$, with $\varepsilon$ bounded by
\begin{equation}
    \varepsilon \leq K \sqrt{2 \mathcal{L}}
    \; .
\end{equation}

\end{lem}
Using this lemma, we have
\begin{equation}
    \mathcal{F}_c\geq 1-2K^2\mathcal{L}
    \; .
\end{equation}
So if $\mathcal{L}=O(\gamma^{t+1})$ then $1-\mathcal{F}_c=O(\gamma^{t+1})$ and thus $1-\mathcal{F}=O(\gamma^{t+1})$.

%%%%%%%%%%%%%%%%%%%%%%%%%%%%%%%%%%%%%%%%%%%%%%%%%%%%%%%%%%%
\section{Recovery and fidelity of $(\!(4,1)\!)$ self-complementary NSA code}\label{Sec:recovery_4_qubit_SC_NSA}

Here we adopt the recovery channel for the non-NSA $(\!(4,1)\!)$ amplitude damping code in~\cite{Leung_1997}, and showcase that the self-complementary NSA code can obtain a larger fidelity using a similar decoding circuit.

We consider the input qubit state,
\begin{equation}
|\bar\psi_{\mathrm{in}}\rangle=a|\bar 0\rangle+b|\bar 1\rangle
\; ,
\end{equation}
where the codewords of the $4$-qubit NSA AD code are as follows:
\begin{align}
\label{eq:412NSA}
    | \bar 0 \rangle& = \left( |0000\rangle + (1-\gamma)^{-2} |1111\rangle \right)/\sqrt{1 + (1-\gamma)^{-4}}\;;\nonumber \\
    | \bar 1 \rangle& = \left( |0011\rangle + |1100\rangle \right)/\sqrt{2}\;.
\end{align}
After $t=1$ amplitude damping errors $A_{0000} = A^0 A^0 A^0 A^0$, $A_{1000} = A^1 A^0 A^0 A^0$, $A_{0100} = A^0 A^1 A^0 A^0$, $A_{0010} = A^0 A^0 A^1 A^0$ and $A_{0001} = A^0 A^0 A^0 A^1$, all possible final states occurring with probabilities at least $O (\gamma)$ are,
\begin{align}
    |\phi_{0000}\rangle&=a\left[\frac{|0000\rangle+|1111\rangle}{\sqrt{1+(1-\gamma)^{-4}}}\right]+b\left[\frac{(1-\gamma)\left[|0011\rangle+|1100\rangle\right]}{\sqrt{2}}\right] 
    \; ,\\
    |\phi_{1000}\rangle&=a\sqrt{\frac{\gamma}{(1-\gamma)(1+(1-\gamma)^{-4})}}|0111\rangle+b\sqrt{\frac{\gamma(1-\gamma)}{2}}|0100\rangle
    \; ,\\
    |\phi_{0100}\rangle&=a\sqrt{\frac{\gamma}{(1-\gamma)(1+(1-\gamma)^{-4})}}|1011\rangle+b\sqrt{\frac{\gamma(1-\gamma)}{2}}|1000\rangle
    \; ,\\
    |\phi_{0010}\rangle&=a\sqrt{\frac{\gamma}{(1-\gamma)(1+(1-\gamma)^{-4})}}|1101\rangle+b\sqrt{\frac{\gamma(1-\gamma)}{2}}|0001\rangle
    \; ,\\
    |\phi_{0001}\rangle&=a\sqrt{\frac{\gamma}{(1-\gamma)(1+(1-\gamma)^{-4})}}|1110\rangle+b\sqrt{\frac{\gamma(1-\gamma)}{2}}|0010\rangle
    \; .
\end{align}

The first step of the recovery process is to perform the error syndrome detection as shown in Fig.~2 in~\cite{Leung_1997}. Based on the syndrome measurement outcomes, one can identify three cases: 1) $A_{0000}$ has occurred, and the state is in $|\phi_{0000}\rangle$; 2) $A_{1000}$ or $A_{0100}$ has occurred, and the state is either in $|\phi_{1000}\rangle$ or $|\phi_{0100}\rangle$; 3) $A_{0010}$ or $A_{0001}$ has occurred, and the state is either in $|\phi_{0010}\rangle$ or $|\phi_{0001}\rangle$.
In the following, we examine these three cases separately.

After the syndrome measurement, only the first the third qubits remain. We denote their states as $|q_1\rangle$ and $|q_3\rangle$ respectively. If $A_{0000}$ has occurred, then $|q_1, q_3\rangle$ after measurement becomes,
\begin{equation}
    a\left[\frac{|00\rangle+|11\rangle}{\sqrt{1+(1-\gamma)^{-4}}}\right]+b\left[\frac{(1-\gamma)(|01\rangle+|10\rangle)}{\sqrt{2}}\right]
    \; .
\end{equation}
Applying a control-not using $|q_3\rangle$ as a control, gives,

\begin{equation}
    |q_1q_3\rangle=\left[(a\sqrt{\frac{2}{1+(1-\gamma)^{-4}}}|0\rangle+b(1-\gamma)|1\rangle)\right]\frac{|0\rangle+\ket{1}}{\sqrt{2}}
    \; .
\end{equation}
We then connect the system with an ancilla qubit prepared to be $|0\rangle$, and consider a non-unitary operation on the $q_1$ qubit: $\mathcal{N}(\rho)=N_0\rho N_0^\dagger+N_1\rho N_1^\dagger$ where
\begin{align}
    N_0&=|0\rangle\langle0|+\sqrt{\frac{2}{(1-\gamma)^2+(1-\gamma)^{-2}}}|1\rangle\langle 1|
    \; ;\\
    N_1&=\sqrt{1-\frac{2}{(1-\gamma)^2+(1-\gamma)^{-2}}}|1\rangle\langle 1|
    \; .
\end{align}
This operation is equivalent to applying a controlled rotation gate $CR_y$—with $\ket{q_1}$ as the control qubit and the ancilla as the target qubit—using a rotation angle of $2\cos^{-1}\left(\sqrt{\frac{2}{(1-\gamma)^2+(1-\gamma)^{-2}}}\right)$. Subsequently, the ancilla qubit is measured. If the measurement outcome on the ancilla is $\ket{0}$, the resulting state is:

\begin{equation}
    |\phi_{\mathrm{out}}\rangle=\sqrt{\frac{2}{1+(1-\gamma)^{-4}}}(a|0\rangle+b|1\rangle)
    \; .
\end{equation}

If $A_{1000}$ or $A_{0100}$ has occurred, then the final states that we should consider will be $|\phi_{1000}\rangle$ or $|\phi_{0100}\rangle$. In either case, $|q_1q_3\rangle$ is a product state and the third qubit will be distorted as
\begin{equation}
    |q_3\rangle=a\sqrt{\frac{\gamma}{(1-\gamma)(1+(1-\gamma)^{-4})}}|1\rangle+b\sqrt{\frac{\gamma(1-\gamma)}{2}}|0\rangle
    \; .
\end{equation}
We can also connect the system with an ancilla qubit prepared to be $\ket{0}$, and considering a non-unitary operation $\mathcal{N}(\rho)=N_0\rho N_0^\dagger+N_1\rho N_1^\dagger$ where
\begin{align}
    N_0&=|0\rangle\langle1|+\sqrt{\frac{2}{(1-\gamma)^2+(1-\gamma)^{-2}}}|1\rangle\langle 0|
    \; ;\\
    N_1&=\sqrt{1-\frac{2}{(1-\gamma)^2+(1-\gamma)^{-2}}}|1\rangle\langle 0|
    \; .
\end{align}
This operation consists of applying an Pauli $X$ gate to qubit $\ket{q_3}$, followed by a controlled rotation gate $CR_y$—with $\ket{q_3}$ as the control qubit and the ancilla as the target qubit—using the same rotation angle as previously described for $A_{0000}$. Subsequently, the ancilla qubit is measured. If the measurement outcome on the ancilla is $\ket{0}$, the resulting state is:

\begin{equation}
    |\phi_{\mathrm{out}}\rangle=\sqrt{\frac{\gamma}{(1-\gamma)+(1-\gamma)^{-3}}}\left[a|0\rangle+b|1\rangle\right]
    \; .
\end{equation}
Same decoding process can be performed when error $A_{0010}$ or $A_{0001}$ has occurred. Thus the worst case fidelity is
\begin{align}
    \mathcal{F}&=\frac{2}{1+(1-\gamma)^{-4}}+\frac{4\gamma}{(1-\gamma)+(1-\gamma)^{-3}}
    =1-3\gamma^2+O(\gamma^3)
    \; .
    \label{eq:nsa_sc_4}
\end{align}
In contrary, the fidelity for the non-NSA $4$-qubit self-complementary AD code in~\cite{Leung_1997} is $1-5\gamma^2+O(\gamma^3)$. Our NSA code indeed has better fidelity at the order of $O(\gamma^2)$.

%%%%%%%%%%%%%%%%%%%%%%%%%%%%%%%%%%%%%%%%%%%%%%%%%%%%%%%%%%%
\section{Recovery and fidelity of $(\!(4,1)\!)$ pair-complementary NSA code}
\label{sec:4-qubit cyclic}
%\shuwei{Considering to mention the loss function of this code}

In this section, we present the recovery protocol and calculate the worst-case fidelity achievable without residual logical errors for the $(\!(4,1)\!)$ pair-complementary NSA code. Consider the following codewords, 
\begin{align}
    \ket{\bar 0}&=\frac{1}{N_0}\left((1-\gamma)^{-1}\ket{0011}-(1-\gamma)^{-3/2}\ket{1110}-(1-\gamma)^{-3/2}\ket{1101}+\ket{0000} 
    \right) \; ; \nonumber \\
    \ket{\bar 1}&=\frac{1}{N_1}\left(
(1-\gamma)^{-1}\ket{1100}+(1-\gamma)^{-1/2}\ket{0001}+(1-\gamma)^{-1/2}\ket{0010}+(1-\gamma)^{-2}\ket{1111}
    \right) \; ,
    \label{eqn:4-qubit cyclic codeword}
\end{align}
where $N_0$ and $N_1$ are normalization factors,
\begin{align}
    N_0=\sqrt{(1-\gamma)^{-2}+2(1-\gamma)^{-3}+1}\; ,  \quad N_1 = \sqrt{(1-\gamma)^{-2}+2(1-\gamma)^{-1}+(1-\gamma)^{-4}}\; .
\end{align}

For an input state $\ket{\bar\psi_{\text{in}}} = a\ket{\bar0} + b\ket{\bar1}$, the possible output states after $t=1$ amplitude damping, occurring with probabilities of order $O(\gamma)$ or higher, are:
\begin{align}
    \ket{\phi_{0000}}&=\frac{a}{N_0}\left(
    \ket{0011}-\ket{1110}-\ket{1101}+\ket{0000} 
    \right) +\frac{b}{N_1}\left(\ket{1100}+\ket{0001}+\ket{0010}+\ket{1111}
    \right)\; ,
    \\
    \ket{\phi_{1000}}&=\sqrt{\frac{\gamma}{1-\gamma}}\left[
    \frac{-a}{N_0}\left(\ket{0110}+\ket{0101}
    \right)+\frac{b}{N_1}\left(
    \ket{0100}+\ket{0111}
    \right)
    \right]\; ,\\
    \ket{\phi_{0100}}&=\sqrt{\frac{\gamma}{1-\gamma}}\left[
    \frac{-a}{N_0}\left(\ket{1010}+\ket{1001}
    \right)+\frac{b}{N_1}\left(
    \ket{1000}+\ket{1011}
    \right)
    \right]\; ,\\
    \ket{\phi_{0010}}&=\sqrt{\frac{\gamma}{1-\gamma}}\left[
    \frac{a}{N_0}\left(\ket{0001}-\ket{1100}
    \right)+\frac{b}{N_1}\left(
    \ket{0000}+\ket{1101}
    \right)
    \right]\; ,\\
    \ket{\phi_{0001}}&=\sqrt{\frac{\gamma}{1-\gamma}}\left[
    \frac{a}{N_0}\left(\ket{0010}-\ket{1100}
    \right)+\frac{b}{N_1}\left(
    \ket{0000}+\ket{1110}
    \right)
    \right] \; . 
    \label{eq:nsa_cc_4}
\end{align}
Note that while orthogonality can assure us to distinguish the errors $A_{0000}, A_{1000}, A_{0100}$, there is a large overlap between the the error words $\ket{\phi_{0010}}$ and $\ket{\phi_{0001}}$, and we cannot distinguish the error $ A_{0010}$ and $A_{0001}$ with a projective measurement. However, one can consider the classical mixture of the two error words and finish the error correction in its diagonal basis.
More explicitly, consider the following projective measurement:
\begin{align}
    P_{0000}=&\frac{1}{4}\left(
    \ket{0011}-\ket{1110}-\ket{1101}+\ket{0000} 
    \right) \left(
    \bra{0011}-\bra{1110}-\bra{1101}+\bra{0000} 
    \right) \nonumber\\
    &+\frac{1}{4}\left(\ket{1100}+\ket{0001}+\ket{0010}+\ket{1111} \right)\left(\bra{1100}+\bra{0001}+\bra{0010}+\bra{1111}
    \right) \; ,   \\
    P_{1000}=&\frac{1}{2}\left(\ket{0110}+\ket{0101}
    \right)\left(\bra{0110}+\bra{0101}
    \right)+\frac{1}{2}\left(
    \ket{0100}+\ket{0111}
    \right)\left(
    \bra{0100}+\bra{0111}
    \right)\; , \\
    P_{0100}=&\frac{1}{2}\left(\ket{1010}+\ket{1001}
    \right)\left(\bra{1010}+\bra{1001}
    \right)+\frac{1}{2}\left(
    \ket{1000}+\ket{1011}
    \right)\left(
    \bra{1000}+\bra{1011}
    \right)\; , \\
    P_S=& \frac{1}{6}\left(\ket{0001}+\ket{0010}-2\ket{1100}
    \right)\left(\bra{0001}+\bra{0010}-2\bra{1100}
    \right)\nonumber \\
    &+ \frac{1}{6}\left(\ket{1101}+\ket{1110}+2\ket{0000}
    \right)\left(\bra{1101}+\bra{1110}+2\bra{0000}
    \right)\; ,
    \label{eqn:S-projection}\\
    P_A=&\frac{1}{2}\left(\ket{0001}-\ket{0010}
    \right)\left(\bra{0001}-\bra{0010}
    \right)+\frac{1}{2}\left(\ket{1101}-\ket{1110}
    \right)\left(\bra{1101}-\bra{1110}
    \right) \; .
    \label{eqn:A-projection}
\end{align}
Here, we use $S$ and $A$ to denote the `symmetrized' and `anti-symmetrized' invariant subspaces. After performing the projection operators, we can further apply some recovery operations based on the projective measurement outcomes. As mentioned before, we choose recovery operations such that there is no residual logical error in the recovered state.

We can now explicitly compute the worst-case fidelity achievable without residual logical errors.
The contribution from the error operator $P_{0000}$ is given by  
\begin{equation}
 \mathcal{F}_0 = 4\min\left\{\frac{1}{N_0^2}, \frac{1}{N_1^2}\right\} = 1-2\gamma+\frac{1}{4}\gamma^2+O(\gamma^3) \;.
\end{equation}
Here, the factor of $4$ arises because the error state $\ket{\phi_{0000}}$ contains four terms.
 The contribution from $P_{1000}$ or $P_{0010}$ is  
 \begin{equation}
     \mathcal{F}_{1}=\frac{2\gamma}{1-\gamma}\min\left\{\frac{1}{N_0^2},\frac{1}{N_1^2}\right\}=\frac{1}{2}\gamma-\frac{1}{2}\gamma^2+O(\gamma^3)\; .
 \end{equation}
Here we use subscript $0$ or $1$ to denote the scaling order with respect to the noise strength $\gamma$ of the code states after the amplitude damping errors.

As $P_{S}$ and $P_{A}$ are the projections to the invariant subspace of the error subspace, the output states, despite being affected by the noise channel, remain in a pure state after projective measurement, namely
\begin{align}
    P_S(\ket{\phi_{0010}}
    \bra{\phi_{0010}}
    +
    \ket{\phi_{0001}}
    \bra{\phi_{0001}}
    )P_S &= \ket{\phi_S}
    \bra{\phi_S}\;,\\
    P_A(\ket{\phi_{0010}}
    \bra{\phi_{0010}}
    +
    \ket{\phi_{0001}}
    \bra{\phi_{0001}}
    )P_A &= \ket{\phi_A}
    \bra{\phi_A}\;,
\end{align}
where the projected states are
\begin{align}           
    \ket{\phi_S}&=\sqrt{\frac{\gamma}{1-\gamma}}\left[
    \frac{a}{\sqrt{2}N_0}(\ket{0001}+\ket{0010}-2\ket{1100})+ \frac{b}{\sqrt{2}N_1}(\ket{1101}+\ket{1110}+2\ket{0000})
    \right]\;,\\
    \ket{\phi_A}&=\sqrt{\frac{\gamma}{1-\gamma}}\left[
    \frac{a}{\sqrt{2}N_0}(\ket{0001}-\ket{0010})+ \frac{b}{\sqrt{2}N_1}(\ket{1101}-\ket{1110})
    \right]\;.
\end{align}
Their contributions to the worst-case fidelity are
\begin{equation}
    \mathcal{F}'_{1}= \mathcal{F}'_{1,S}+ \mathcal{F}'_{1,A}=(3+1)\frac{\gamma}{1-\gamma}\min\left\{\frac{1}{N_0^2},\frac{1}{N_1^2}\right\}=\gamma-\gamma^2+O(\gamma^3)\;.
\end{equation}
To summarize, we get the worst-case fidelity up to $O(\gamma^3)$,
\begin{equation}
    \mathcal{F} = \mathcal{F}_0 + 2 \mathcal{F}_{1}+ \mathcal{F}'_{1}= 1- \frac{7}{4} \gamma^2+O(\gamma^3)
    \; ,
\end{equation}
which is higher than the worst-case fidelity Eq.~(\ref{eq:nsa_sc_4}) of the $(\!(4,1)\!)$ self-complementary NSA code.

%%%%%%%%%%%%%%%%%%%%%%%%%%%%%%%%%%%%%%%%%%%%%%%%%%%%%%%%%%%
\section{NSA Self-Complementary (SC) Qubit Code Generalization}

In this section, we show how to generalize the $(\!(4,1)\!)$ NSA AD code to a family of self-complementary (SC) NSA AP codes, and compare the fidelity difference between the NSA and non-NSA AD codes for arbitrary $n$.

As shown in~\cite{lang2007,QuditAD_Zeng2009}, for a SC $(\!(n,k)\!)$ code $C=\operatorname{span}\{(|u\rangle+|\tilde{u}\rangle) / \sqrt{2} \mid u \in S\}$ can correct $t=1$ amplitude damping errors if and only if the set $S \subset\{0,1\}^n$ satisfies: 1) if $u \in S$, then $\bar{u} \in S$; 2) error spaces from different $u$ must not overlap. Note that we consider largest possible $k$ given a $n$ in the SC for the following calculations. A non-NSA SC AD code can be written as,
\begin{equation}
    | \bar{\psi}_u  \rangle = \frac{1}{\sqrt 2} \left( | u \rangle + | \tilde u \rangle \right)
    \; ,
\end{equation}
where $\tilde u_i = 1 - u_i$ for every single qubit site. The AD error that has $A_0$ acting on every single qubit transforms an input state $| \bar\psi_{\text{in}} \rangle = \sum_u c_u  \left( | u \rangle + | \tilde u \rangle \right) / \sqrt{2}$ into 
\begin{equation}
    \sum_a c_u \left(\sqrt{\frac{(1-\gamma)^{\|u\|} }{2}}  | u \rangle +  \sqrt{\frac{ (1-\gamma)^{\| \tilde u \|}}{2}}| \tilde u \rangle \right) 
    \; ,
\end{equation}
where $\| u \|$ denotes the Hamming weight of the code word basis $\ket u$. The contribution to the worst-case fidelity is then dominated by the case where the number of 0s and 1s in the bit string $u$ is balanced. Thus
\begin{equation}
    \mathcal{F}^{\text{non-NSA}}_0= 
        \frac{(1-\gamma)^{\lfloor n/2 \rfloor}+(1-\gamma)^{n-\lfloor n/2 \rfloor}}{2} \;.
\end{equation}

Consider an AD error that has $A^1$ acting only on one qubit $i$ and $A^0$ acting on the rest of the qubits, denoted as $A_{1_i}$. It transforms the input state into,
\begin{equation}
\label{eq:SC_nonNSA_afterA1}
    \frac{1}{\sqrt{2}} \left(
        \delta_{1,u_i} (1-\gamma)^{(\|u\| - 1)/2} \gamma^{1/2} | u - e_i \rangle 
        + \delta_{1,\tilde u_i} (1-\gamma)^{(\|\tilde u\| -1)/2} \gamma^{1/2} | \tilde u - e_i \rangle
    \right)
    \; ,
\end{equation}
where $e_i$ denotes a bit-string with 1 at the i-th bit and 0 on the others.
Note that after the $A^1$ noise, only one of the two bases in Eq.~(\ref{eq:SC_nonNSA_afterA1}) survives, whose weight takes the minimum value when $u$ contains all zeros or all ones. This gives the contribution of the worst-case fidelity
\begin{equation}
    \mathcal{F}^{\text{non-NSA}}_1=\frac{ \, (1-\gamma)^{(n-1)} \, \gamma }{2}\;.
\end{equation}
Hence, the worst-case fidelity is,
\begin{equation}
    \mathcal{F}^{\text{non-NSA}}
    = 
       \mathcal{F}^{\text{non-NSA}}_0+n\mathcal{F}^{\text{non-NSA}}_1
    = 
        1 - (n^2 - n + 2n\lfloor n/2 \rfloor - 2 \lfloor n/2 \rfloor^2) \gamma^2 / 4  + \mathcal{O}(\gamma^3)
    \; .
\end{equation}

Now, we generalize the non-NSA SC codeword to the following NSA SC codeword for arbitrary $n$,
\begin{equation}
    | \bar \psi_u \rangle = \frac{
    (1 - \gamma)^{ - \|u\|/2}|u \rangle 
    + (1 - \gamma)^{ - \|\tilde u\|/2}|\bar u \rangle}
    {\sqrt{(1 - \gamma)^{-\|u\|} + (1 - \gamma)^{ - \|\tilde u\|}}} 
    \; .
\end{equation}
Similarly, consider an input state that is a random superposition of all logical qubits. The AD error $A_0$ transforms the input state into,
\begin{equation}
    \sum_u c_u \frac{|u \rangle + |\tilde u \rangle}{\sqrt{(1-\gamma)^{-\|u\|}+(1-\gamma)^{-\|\tilde u\|}} }
    \; .
\end{equation}
Its contribution to the worst-case fidelity is 
\begin{equation}
    \mathcal{F}_0^{\text{NSA}} = \min_u \left\{
    \frac{2}{(1-\gamma)^{-\|u\|}+(1-\gamma)^{-(n-\|u\|)}}
    \right\}= \frac{2}{1+(1-\gamma)^{-n}}
\end{equation}
Using the mean inequality, we find that when $u$ is the all-zero bit string (or equivalently, the all-one bit string), the minimum can be achieved.

And the AD error $A_{1_i}$ transforms the input state into,
\begin{equation}
   \sum_u c_u \, (1-\gamma)^{-1/2} \, \gamma^{1/2} \, 
   \frac{\delta_{1, u_i}  | u - e_i \rangle 
        + \delta_{1,\tilde u_i} | \tilde u - e_i \rangle }{\sqrt{(1 - \gamma)^{-\|u\|} + (1 - \gamma)^{ - \|\tilde u\|}}} \;
    \; .
\end{equation}
Its contribution to the worst-case fidelity is 
\begin{equation}
    \mathcal{F}_1^{\text{NSA}} = \min_u \left\{ \frac{\gamma\, (1-\gamma)^{-1}}{(1-\gamma)^{-\|u\|}+(1-\gamma)^{-(n-\|u\|)}}\right\} = \frac{\gamma}{(1-\gamma)^{1-n}+(1-\gamma)}
\end{equation}
As above, using the mean inequality, we find that when $u$ contains all zeros or all ones, the minimum can be achieved.

The total contribution to the worst-case fidelity hence is, 
\begin{equation}
    \mathcal{F}^{\text{NSA}} = 
        \mathcal{F}_0^{\text{NSA}} 
        +n \mathcal{F}_1^{\text{NSA}} 
    = 1 - (n^2-n) \, \gamma^2 / 4 + \mathcal{O}(\gamma^3)
    \; .
    \label{eq:f_nsa_sc_qubit}
\end{equation}
The fidelity for the NSA SC codes is shown to be better than the non-NSA SC codes, and their difference is $(2n\lfloor n/2 \rfloor - 2 \lfloor n/2 \rfloor^2)/4$ in the order of $\gamma^2$. Notice that the advantage of NSA SC code grows as the qubit number $n$ increases.

%%%%%%%%%%%%%%%%%%%%%%%%%%%%%%%%%%%%%%%%%%%%%%%%%%%%%%%%%%%
\section{NSA Self-Complementary Qudit Code Generalization}

In this section, we discuss the generalization to the NSA SC {\it qudit} code with local dimension $q \ge 2$. 
We start with an example of $n=4$ qutrit codes. Then, we show the definition of the amplitude damping noise for qudit systems, and detail the extension to the non-NSA SC qudit codes from the non-NSA SC qubit codes. Afterwards, we show how we can generalize the non-NSA SC qudit codes to the NSA SC qudit codes. Along the way, we show how to compute the worst-case fidelities for both non-NSA and NSA SC qudit codes, and argue that NSA SC qudit codes have better performance. 

We start with an example of $n=4$, $q=3$ amplitude damping code. We first introduce the non-NSA version of the code and then move to the NSA version, and calculate their fidelities respectively. For $q=3$, the amplitude damping operator takes the form, 
\begin{align}
    A_0 &= | 0 \rangle \langle 0 | + \sqrt{1-\gamma} |1 \rangle \langle 1 | + (1-\gamma) |2 \rangle \langle 2|
    \; , \\
    A_1 &=  \sqrt{\gamma} |0 \rangle \langle 1 | + \sqrt{2\gamma(1-\gamma)} |1 \rangle \langle 2|
    \; , \\
    A_2 &= \gamma |0 \rangle \langle 2|
    \; .
\end{align}
The non-NSA codewords are the following,
\begin{align}
    | \bar{0} \rangle
		&=
		\sqrt{\frac{1}{3}}(|0000\rangle+|1111\rangle+|2222\rangle)
	\; ;
	\nonumber
	\\
    | \bar{1} \rangle
		&=
		\sqrt{\frac{1}{3}}(|0011\rangle+|1122\rangle+|2200\rangle)
    \; ;
	\nonumber
	\\
    | \bar{2} \rangle
		&=
		\sqrt{\frac{1}{3}}(|0022\rangle+|1100\rangle+|2211\rangle)     
	\; .
\end{align}
Consider an input state $|\bar \psi\rangle = a |\bar 0 \rangle + b |\bar 1 \rangle + c |\bar 2 \rangle$. The output states after error operators $A_{0000}$, $A_{1000}$ and $A_{2000}$ respectively are,
\begin{align}
    |\psi_{0000}\rangle 
    &= a\left[\frac{|0000\rangle+(1-\gamma)^2|1111\rangle+(1-\gamma)^4|2222\rangle}{\sqrt{3}}\right]\nonumber \\
    &+b\left[\frac{\left[(1-\gamma)|0011\rangle+(1-\gamma)^3|1122\rangle+(1-\gamma)^2|2200\rangle\right]}{\sqrt{3}}\right]\nonumber \\
    &+c\left[\frac{\left[(1-\gamma)^2|0022\rangle+(1-\gamma)|1100\rangle+(1-\gamma)^3|2211\rangle\right]}{\sqrt{3}}\right]
    \; ; \\
    |\psi_{1000}\rangle 
    &= a\left[\frac{\sqrt{\gamma}(1-\gamma)^{3/2}|0111\rangle+\sqrt{2\gamma}(1-\gamma)^{7/2}|1222\rangle}{\sqrt{3}}\right]\nonumber \\
    &+b\left[\frac{\sqrt{\gamma}(1-\gamma)^{5/2}|0122\rangle+\sqrt{2\gamma}(1-\gamma)^{3/2}|1200\rangle}{\sqrt{3}}\right]\nonumber \\
    &+c\left[\frac{\sqrt{\gamma}(1-\gamma)^{1/2}|0100\rangle+\sqrt{2\gamma}(1-\gamma)^{5/2}|1211\rangle}{\sqrt{3}}\right]
    \; ; \\
    |\psi_{2000}\rangle 
    &= a\left[\frac{\gamma(1-\gamma)^{3}|0222\rangle}{\sqrt{3}}\right]\nonumber \\
    &+b\left[\frac{\gamma(1-\gamma)|0200\rangle}{\sqrt{3}}\right]\nonumber \\
    &+c\left[\frac{\gamma(1-\gamma)^{2}|0211\rangle}{\sqrt{3}}\right]
    \; .
\end{align}
Note that there is no overlap between the error subspaces, enabling us to distinguish different errors by a projective measurement. The error correction procedure is almost identical to section~\ref{Sec:recovery_4_qubit_SC_NSA}, and the total worse-case fidelity would be 
\begin{align}
    \mathcal{F}^{\text{non-NSA SC}}_{\text{4\_qutrit}} &= ((1 - \gamma)^2 + (1 - \gamma)^4 + (1 - \gamma)^6)/3 + 
    4 \times (\gamma (1 - \gamma)^3 + 2 \gamma (1 - \gamma)^7)/3 + 4 \times (\gamma^2 (1 - \gamma)^6)/3 \nonumber \\
    &=1-14\gamma^2+O(\gamma^3)
    \; .
\end{align}

Now, we move on to NSA case. Inspired by the qubit case, we can choose the logical states to be the following,
\begin{align}
    |\bar0 \rangle &= \frac{ | 0000 \rangle + (1-\gamma)^{-2} | 1111 \rangle + (1-\gamma)^{-4}| 2222 \rangle}
    {\sqrt{1 + (1-\gamma)^{-4} + (1-\gamma)^{-8}}} \; ;
    \nonumber \\
    |\bar1 \rangle &= \frac{(1-\gamma)^{-1} | 0011 \rangle + (1-\gamma)^{-3} | 1122 \rangle + (1-\gamma)^{-2} | 2200 \rangle}
    {\sqrt{(1-\gamma)^{-2} + (1-\gamma)^{-6} + (1-\gamma)^{-4}}} \; ;
    \nonumber \\
    |\bar2 \rangle &= \frac{ (1-\gamma)^{-2} | 0022 \rangle+(1-\gamma)^{-1} | 1100 \rangle 
    + (1-\gamma)^{-3} | 2211 \rangle 
    }
    {\sqrt{(1-\gamma)^{-2} + (1-\gamma)^{-6} + (1-\gamma)^{-4}}}
    \; .
\end{align}

After affected by noise $A_{0000}, ~A_{1000}, ~A_{2000}$, 
\begin{align}
    | \phi_{0000} \rangle = 
    &a_0 \left[
    \frac{ | 0000 \rangle +  | 1111 \rangle + | 2222 \rangle}
    {\sqrt{1 + (1-\gamma)^{-4} + (1-\gamma)^{-8}}}
    \right] \nonumber \\
    + &a_1 \left[
    \frac{ | 0011 \rangle +  | 1122 \rangle + | 2200 \rangle}
    {\sqrt{(1-\gamma)^{-2} + (1-\gamma)^{-6} + (1-\gamma)^{-4}}}
    \right] \nonumber \\
    + &a_2 \left[
    \frac{ | 0022 \rangle +  | 1100 \rangle +  | 2211 \rangle}
    {\sqrt{(1-\gamma)^{-2} + (1-\gamma)^{-6} + (1-\gamma)^{-4}}}
    \right]
    \; ; \\
    | \phi_{1000} \rangle =
    &a_0 \left[
    \frac{\gamma^{1/2}| 0111 \rangle + (2\gamma)^{1/2} | 1222 \rangle}
    {\sqrt{(1-\gamma) + (1-\gamma)^{-3} + (1-\gamma)^{-7}}}
    \right] \nonumber\\
    + &a_1 \left[ 
    \frac{\gamma^{1/2} | 0122 \rangle + (2\gamma)^{1/2}| 1200 \rangle}
    {\sqrt{(1-\gamma)^{-1} + (1-\gamma)^{-5} + (1-\gamma)^{-3}}}
    \right]\nonumber\\
    + &a_2 \left[ 
    \frac{\gamma^{1/2} | 0100 \rangle + (2\gamma)^{1/2} | 1211 \rangle}
    {\sqrt{(1-\gamma)^{-1} + (1-\gamma)^{-5} + (1-\gamma)^{-3}}}
    \right]
    \; ; \\
    | \phi_{2000} \rangle =
    &a_0 \left[ 
    \frac{\gamma\, | 0222 \rangle}
    {\sqrt{(1-\gamma)^2 + (1-\gamma)^{-2} + (1-\gamma)^{-6}}}
    \right]\nonumber\\
    + &a_1 \left[ 
    \frac{\gamma \, | 0200 \rangle}
    {\sqrt{1 + (1-\gamma)^{-4} + (1-\gamma)^{-2}}}
    \right] \nonumber\\
    &a_2 \left[ 
    \frac{\gamma\, | 0211 \rangle}
    {\sqrt{(1 + (1-\gamma)^{-4} + (1-\gamma)^{-2}}}
    \right]
    \; .
\end{align}
The error correction procedure is also very similar to section~\ref{Sec:recovery_4_qubit_SC_NSA}, and the total worse-case fidelity would be 
\begin{align}
    \mathcal{F}^{\text{NSA SC}}_{\text{4\_qutrit}} 
    &= \frac{3}{1+(1-\gamma)^{-4}+(1-\gamma)^{-8}}
    + 4 \times \frac{3\gamma}{(1-\gamma) + (1-\gamma)^{-3} + (1-\gamma)^{-7}} 
    + 4 \times \frac{\gamma^2 }{(1-\gamma)^2 + (1-\gamma)^{-2} + (1-\gamma)^{-6}} \nonumber \\
    &= 1 - 10 \gamma^2 + \mathcal{O} (\gamma^3) \; .
\end{align}

Now, consider the amplitude damping noises for general qudit systems that take the following form,
\begin{equation}
    A^l = \sum_{a=l}^{q-1} \sqrt{\binom{a}{l}} \, \sqrt{(1-\gamma)^{a-l} \gamma^l} \, | a-l \rangle \langle a |
    \; ,
\end{equation}
where $l$ indicates the level of jumps in the AD noises. 

We move to the general qudit case for both the non-NSA and NSA codes. To demonstrate the structure, we first introduce the non-NSA version. Inspired by the non-NSA SC qubit codes, we define a new family of non-NSA SC qudit codes with local dimension $q$ as $\mathfrak{C}_{d} = \operatorname{span}\{ \sum_{a=0}^{q-1} \ket{u^{\uparrow_a}}/ \sqrt{q} | u \in S_d \}$ for $S_d \subset \{0, \cdots , (q-1)\}^n$, where $ \ket{u^{\uparrow_a}} = \ket{( u + a^{\otimes n} ) \operatorname{mod} q} $ and the module is taken at every qudit site. 
Here, we continue to consider weight $t=1$ amplitude damping errors, where there is at most one non-$A_0$ acting on all qudits. 
Code $\mathfrak{C}_{d}$ can correct $t=1$ AD errors if and only if the set $S_d$ satisfies: 1) if $u \in S_d$, then $ u^{\uparrow_a} \in S_d,~ \forall \, a \in \{0,\cdots,(q-1)\}$; 2) error spaces from different $u$ must not overlap~\cite{lang2007, Kong2015}.
%\shiyu{I realized that we never proved it...}
The codeword takes the form of, 
\begin{equation}
    | \bar{\psi}_u \rangle
        =\frac{1}{\sqrt{q}}    
            \sum_{a=0}^{q-1} \, \ket{u^{\uparrow_a}}
    \; .
    \label{eq:qudit_cw}
\end{equation}

Notice that $a$ enumerates each basis in the codeword. Consider the following input state,
\begin{equation}
    |\bar\psi_\text{in} \rangle = \frac{1}{\sqrt{q}}\sum_u c_u \sum_{a=0}^{q-1} | u^{\uparrow_a} \rangle
    \; .
\end{equation}
An amplitude damping error with only $A^0$ acting on every single qudit changes the input state to,
\begin{equation}
    \frac{1}{\sqrt{q}}\sum_u c_u 
    \sum_{a=0}^{q-1} (1-\gamma)^{\|u^{\uparrow_a}\|/2}| u^{\uparrow_a} \rangle
    \; .
\end{equation}
Its contribution to the worst-case fidelity is taken from the codeword that has the smallest coefficients. 
For error of all $A_0$, the codeword that has the smallest coefficient is the one where $u^{\uparrow_0}$ is half being $0$ and the rest being $1$.
\begin{equation}
    \min_u \left\{ \frac{ \sum_{a=0}^{q-1} (1-\gamma)^{\|u^{\uparrow_a}\|}}{q} \right\} 
    = \min_u\left[1-\frac{1}{q}\left(\sum_{a=0}^{q-1} \|u^{\uparrow_a} \|\right)\gamma +\frac{1}{2q} \left(\sum_{a=0}^{q-1}\left(-\|u^{\uparrow_a} \|+\|u^{\uparrow_a} \|^2\right) \right)\gamma^2  + \cdots
    \right]\; .
    \label{eqn:SC-non-NSA-F0}
\end{equation}
Notice that every dit in $u^{\uparrow_a}$ is a permutation of $\{0,1,...,q-1\}$ as $a$ runs from 0 to $q-1$, hence the summation of weight is a constant
\begin{equation}
    \sum_a\|u^{\uparrow_a}\|=nq(q-1)/2 \; .
\end{equation}
Thus, finding the minimum of Eq.\eqref{eqn:SC-non-NSA-F0} is equivalent to finding the minimum of $\sum_{a=0}^{q-1}\|u^{\uparrow_a}\|^2$. For simplicity of the calculation, we assume that $n$ is divisible by $q$. By the mean inequality we can minimize the summation of the square by taking $\|u^{\uparrow_a}\|$ to be independent of $a$, which can be achieved by taking $u$ with each of the digits from 0 to $q-1$ appearing $n/q$ times, thus $\|u^{\uparrow_a}\|=n(q-1)/2$, and we obtain the minimum to be 
\begin{equation}
    \min_u\sum_a\|u^{\uparrow_a}\|^2=n^2q(q-1)^2/4 \; .
\end{equation}
By substituting the minimum quantity into Eq.~(\ref{eqn:SC-non-NSA-F0}), we obtain its contribution to the fidelity,
\begin{equation}
       \mathcal{F}_0 = 1 
        - \frac{1}{2}n(q-1) \gamma 
        + \frac{1}{8}n (q-1) (n(q-1)-2)\gamma^2+O(\gamma^3)\;.
\end{equation}

Let's now consider the case where an amplitude damping error that evolves jump of $|u^{\uparrow_a} -e^{(l)}_i \rangle \langle u^{\uparrow_a}|$ where $u^{\uparrow_a}_i \ge l$ denotes the $i$-th dit in the dit string $u^{\uparrow_a}$. We define $e^{(l)}_i$ to be a dit string that is all $0$ except that the $i$-th dit is of value $l$. It effectively represent the AD error which has $A^l$ is acting on qudit site $i$ and $A^0$ acting on all the rest qudits. After the error, the input state becomes,
\begin{equation}
    \frac{1}{\sqrt{q}} \sum_u c_u
    \sum_{a=0}^{q-1} 
    \sqrt{\binom{u^{\uparrow_a}_i}{l}} \, (1-\gamma)^{(\|u^{\uparrow_a}\|-l)/2} \gamma^{l/2}\,
    | u^{\uparrow_a} - e^{(l)}_i\rangle
    \; .
\end{equation}
Since we assumed that different error spaces do not overlap, we can use projective measurement to distinguish the noise state after any $e^{(l)}_i$. The contribution to the worst-case fidelity comes from a codeword of smallest coefficient, and is,
\begin{equation}
    \min_u \left\{ \frac{1}{q}\sum_{a=0}^{q-1} 
    \binom{u^{\uparrow_a}_i}{l} \, (1-\gamma)^{(\|u^{\uparrow_a}\|-l)} \gamma^{l} \right\} 
    \; .
\end{equation}
In the case where $l=1$, 
\begin{align}
    \mathcal{F}_1&=\min_u \left\{ \frac{1}{q}\sum_{a=0}^{q-1} 
    u^{\uparrow_a}_i \, (1-\gamma)^{(\|u^{\uparrow_a}\|-1)} \gamma \right\} 
    = \min_u \left\{ \frac{1}{q}\sum_{a=0}^{q-1}u^{\uparrow_a}_i\gamma-\frac{1}{q}\max_u \sum_{a=0}^{q-1}u^{\uparrow_a}_i(\|u^{\uparrow_a}\| -1)\gamma^2+O(\gamma^3) \right\} \\
    &=\frac{q-1}{2}(\gamma+\gamma^2)-\frac{\gamma^2}{q}
    \max_u \sum_{a=0}^{q-1}u^{\uparrow_a}_i\|u^{\uparrow_a}\| +O(\gamma^3)
    \; .
\end{align}
the minimization is taken when $\sum_{a=0}^{q-1}u^{\uparrow_a}_i\|u^{\uparrow_a}\|$ is taken to be maximal. Note that
\begin{equation}
    \sum_{a=0}^{q-1}u^{\uparrow_a}_i\|u^{\uparrow_a}\|=\sum_j\sum_{a=0}^{q-1}u^{\uparrow_a}_i u^{\uparrow_a}_j\leq n\sum_{a=0}^{q-1} a^2=n\frac{(q-1)q(2q-1)}{6} \;.
\end{equation}
where by the rearrangement inequality, the maximum is taken when $u^{\uparrow_a}_i= u^{\uparrow_a}_j$ for any $i$, $j$ and $a$, i.e. $u$ is a dit string where all dits are the same. Hence,
\begin{equation}
    \mathcal{F}_1= \frac{q-1}{2}\gamma
+\frac{(q-1)\left(3-n(2q-1)\right)}{6}\gamma^2+O(\gamma^3) \;.
\end{equation}

In the case where $l=2$,
\begin{equation}
    \mathcal{F}_2=\sum_{a=2}^{q-1}{a\choose 2}\gamma^2=\frac{q(q-1)(q-2)}{6}\gamma^2+O(\gamma^3) \;.
\end{equation}
where we get rid of the dependence of $u$ because when we sum over $a$, $u^{\uparrow_a}_i$ is just a permutation of integers from 0 to $q-1$. Moreover, note that for $l>2$, its contribution to the worst-case fidelity is already on the order of $O(\gamma^3)$; thus, these terms can be safely neglected.

The  total worst-case fidelity for non-NSA code is
\begin{equation}
    \mathcal F^{\text{non-NSA}} =\mathcal F_0+n\mathcal F_1+n\mathcal F_2  = 1 - \frac{1}{24}\left( 
    n (q-1) (2 - 4q + n (5q - 1))
    \right)\gamma^2+O(\gamma^3) \;.
\end{equation}

For the generalization to NSA qudit amplitude damping code, we follow the similar analysis above to calculate the worst case fidelity. The general form of the NSA codeword with local dimension $q$ is,
\begin{equation}
    | w \rangle 
        =  \frac{\sum_{a=0}^{q-1} \, (1-\gamma)^{(-\|u^{\uparrow_a}\|)/2}}
        {\sqrt{\sum_{a=0}^{q-1} (1-\gamma)^{-\|u^{\uparrow_a}\|}}}
    \, | u^{\uparrow_a} \rangle
    \; ,
\end{equation}
For error with only the damping $A^0$, a general input state becomes,
\begin{equation}
    \sum_u c_u
        \frac{ | u^{\uparrow_a} \rangle}
        {\sqrt{\sum_{a=0}^{q-1} (1-\gamma)^{-\|u^{\uparrow_a}\|}}}
    \; ,
\end{equation}
and contribution to the worst case fidelity is the minimal coefficient of the codeword,
\begin{equation}
    \min_u 
    \left\{ 
        \frac{q}
        {\sum_{a=0}^{q-1} (1-\gamma)^{-\|u^{\uparrow_a}\|}}
    \right\} 
    = 
        \frac{ q }
        {\sum_{a=0}^{q-1} (1-\gamma)^{-a \cdot n}}
    \; .
\end{equation}
Now consider the amplitude damping errors evolving jumps $l$. The input state becomes,
\begin{equation}
    \sum_u c_u
        \frac{\sum_{a=0}^{q-1}     \sqrt{\binom{u^{\uparrow_a}_i}{l}} \, \gamma^{l/2} \, (1-\gamma)^{- l/2}}
        {\sqrt{\sum_{a=0}^{q-1} (1-\gamma)^{-\|u^{\uparrow_a}\|} }}
    \, | u^{\uparrow_a} - e^{(l)}_i \rangle
    \; .
\end{equation}
Its contribution to the worst case fidelity is, 
\begin{equation}
    \min_u 
    \left\{
        \frac{\sum_{a=0}^{q-1}     \binom{u^{\uparrow_a}_i}{l} \, \gamma^{l} \, (1-\gamma)^{- l}}
        {\sum_{a=0}^{q-1} (1-\gamma)^{-\|u^{\uparrow_a}\|}}
    \right\}
    = \frac{\sum_{a=0}^{q-1}     \binom{a}{l} \, \gamma^{l} \, (1-\gamma)^{- l}}{\max_u \sum_{a=0}^{q-1} (1-\gamma)^{-\|u^{\uparrow_a}\|}}=
        \frac{\sum_{a=0}^{q-1}     \binom{a}{l} \, \gamma^{l} \, (1-\gamma)^{- l}}
        {\sum_{a=0}^{q-1} (1-\gamma)^{-a \cdot n}}
    \; .
\end{equation}
From the rearrangement inequality, the smallest coefficient here corresponds to the codeword in which $u$ is all $0$. And the total worst case fidelity is,
\begin{align}
    \mathcal{F}^{\text{NSA}} 
    &= 
        \frac{ q }
        {\sum_{a=0}^{q-1} (1-\gamma)^{-a \cdot n}}
    +
        n \times \sum_{l=1}^{q-1} \;
        \frac{\sum_{a=0}^{q-1}     \binom{a}{l} \, \gamma^{l} \, (1-\gamma)^{ - l}}
        {\sum_{a=0}^{q-1} (1-\gamma)^{-a \cdot n}}
    \nonumber \\
    &= 
        1-\frac{1}{12}(q-1)(2q-1)(n^2-n)\gamma^2+O(\gamma^3)
    \; .
\end{align}

Hence, we find the difference between NSA and non-NSA fidelities, 
\begin{equation}
    \Delta \mathcal F = \frac{1}{24}n^2(q^2-1)\gamma^2
\end{equation}

Below we display fidelities up to second order in $\gamma$ for several different $q$ of general $n$ for both NSA and non-NSA self-complementary codes.

\begin{center}
\begin{tabular}{ | c | c | c | } 
  \hline
  $q$ & $F^{\text{NSA SC}}$ & $F^{\text{non-NSA SC}}$    
  \\ \hline
  3 & $1 - \frac{5}{6}(n^2-n)\gamma^2$ & $1 - \frac{1}{12}(14n^2 - 10n)\gamma^2$
  \\ \hline
  4 & $1 - \frac{7}{4}(n^2-n)\gamma^2$ & $1 - \frac{1}{8}(19 n^2 - 14 n)\gamma^2$  
  \\ \hline
  5 & $1 - 3(n^2-n)\gamma^2$ & $1 - \frac{1}{6}(24 n^2 - 18 n)\gamma^2$  
  \\ \hline
  6 & $1 - \frac{55}{12}(n^2-n)\gamma^2$ & $1 - \frac{5}{24}(29 n^2 - 22 n)\gamma^2$ 
  \\ \hline
  7 & $1 - \frac{13}{2}(n^2-n)\gamma^2$ & $1 - \frac{1}{4}(34 n^2 - 26 n)\gamma^2$ 
  \\ \hline
\end{tabular}
\end{center}

%%%%%%%%%%%%%%%%%%%%%%%%%%%%%%%%%5
\section{NSA pair-complementary qubit code generalization}
In this section, we generalize the $(\!(4,1)\!)$ pair-complementary NSA code to a family of pair-complementary NSA codes, and calculate the worst-case fidelity achievable without residual logical errors for arbitrary $n$.

By inspecting the codewords of the $(\!(4,1)\!)$ pair-complementary NSA code, one observes that the basis states are paired with each other. 
Thus we consider the codewords of the following form:
\begin{align}
    \ket{\bar{\psi}^{i,j}_u}   &\sim |u\rangle+|u-e_i-e_j\rangle-|\tilde{u}+e_i\rangle-|\tilde{u}+e_j\rangle\; ; \nonumber  \\
    \ket{\bar{\psi}'^{i,j}_u} &\sim |u-e_i\rangle+|u-e_j\rangle+|\tilde{u}+e_i+e_j\rangle+|\tilde{u}\rangle\; ,
\end{align} 
where $e_i\in\{0,1\}^n$ denotes the error vector with $1$ at the $i$-th bit and $0$ on the others.
To highlight the structure of this code more clearly, we have omitted the coefficients in the equations above. Specifically, each computational basis state $\ket{u}$ should include a factor of $(1-\gamma)^{-\|u\|/2}$ before it, where $\|u\|$ denotes the Hamming weight of the bit-string $u$.

We then try to find the set of $\{u,i,j\}$ such that the Knill-Laflamme condition can be approximately satisfied. To simplify notation, we define $S_{u,i,j}=\{u,u-e_i,u-e_j,u-e_i-e_j\}$ in the following, and use $\tilde{S}_{u,i,j}=\{\tilde{u},\tilde{u}+e_i,\tilde{u}+e_j,\tilde{u}+e_i+e_j\}$ to denote its element-wise complement.
The pairwise orthogonality condition of codewords then  imposes the following constraint:

(C1) $(S_{u,i,j}\cup\tilde{S}_{u,i,j})\cap(S_{v,k,l}\cup\tilde{S}_{v,k,l}) \neq \varnothing$, iff $u = v$ and $\{i,j\} = \{k,l\}$.

To assure that no confusion between different $\{u,i,j\}$ arise after one damping, we will also require

(C2) Error spaces from different $\{u,i,j\}$ must not overlap.

Since for each triplet $\{u,i,j\}$, the corresponding pair-complementary code words $\ket{\bar\psi^{i,j}_u}$ and $\ket{\bar\psi'^{i,j}_u}$ contains superpositions of all combinations of 0 and 1s on the $i$-th and $j$-th qubits, for our condition (C2) it is only relevant to consider code word on the other $n-2$ qubits. Similarily, for another triplet $\{v,k,l\}$ it is only relevant to consider code word on $n-2$ qubits other than the $k$-th and $l$-th qubits. To maximize the code rate, it is then natural to pick our code spaces such that for any triplet,  $\{u,i,j\}$, the set $\{i,j\}$ are the same. This inspires us to construct the following family of pair-complementary NSA codes.
\begin{prop}
    For any $(\!(n,k)\!)$ self-complementary code \begin{equation}
        \mathfrak{C}_1=\mathrm{span}\{(|u\rangle+ |\tilde{u}\rangle)/\sqrt{2}\ |\ u\in S\}\;,
    \end{equation}
    we can construct a $(\!(n+2,k+1)\!)$ pair-complementary code \begin{equation} \mathfrak{C}_2=\mathrm{span}\{\ket{\bar{\psi}_u},\ket{\bar{\psi}'_{u}}\}\ |\ u\in S/\mathbb{Z}_2\}\;,
    \end{equation}
    where
    \begin{align}
        |\bar\psi_u\rangle&\sim |u00\rangle+|u11\rangle-|\tilde{u}10\rangle-|\tilde{u}01\rangle \nonumber \; ;\\
        |\bar{\psi}'_{u}\rangle&\sim |u01\rangle+|u10\rangle+|\tilde{u}11\rangle+|\tilde{u}00\rangle\; .
        \label{eqn:cyclic-self complementary codeword}
    \end{align}
    Here the quotient set $S/\mathbb{Z}_2$ is defined such that if $u\in S/\mathbb{Z}_2\subset S$ then $\tilde{u}\notin S/\mathbb{Z}_2$, i.e. the quotient set only contains one element in each complementary pairs $u$ and $\tilde{u}$.  $u00$ denotes a bit string starting with $u$ and ending with $00$, with similar conventions for the other terms.
\end{prop}

\begin{proof}
    To show that our construction gives an AQEC code for one amplitude damping error, we verify that it satisfies the condition C1 and C2, and demonstrate the error correction process and the worst case fidelity for it.

    To verify C1, notice that for $S_{u,i,j}\cup\tilde{S}_{u,i,j}$, the first $n-2$ bits in the bit strings are either $u$ or $\tilde{u}$, thus $(S_{u,i,j}\cup\tilde{S}_{u,i,j})\cap(S_{v,k,l}\cup\tilde{S}_{v,k,l})=\varnothing$ iff $u=v$ or $u=\tilde{v}$. Since $u,v\in S/\mathbb{Z}_2$, $\tilde v\notin S/\mathbb{Z}_2$, so $u=v$.

    To verify C2, consider code states generated from different bit strings $u$ and $v$, notice that if the errors happen in the last 2 qubits, then since $u\neq v$ and $u\neq \tilde v$, the error spaces are guaranteed to not overlap with each other. If the errors do not happen in the last 2 qubits, then the non-overlap requirement is equivalent to saying that for $u,\tilde{u},v,\tilde{v}$, there is no confusion arises assuming the decay occurs at no more than one qubit.
    
    This is exactly the condition for $\mathfrak{C}_1$ to be an AQEC code for a single AD noise.
\end{proof}

It is worth mentioning that when we take  $u=00$, the above codeword corresponds to the $(\!(4,1)\!)$ NSA pair-complementary code in Eq.~(5) of the main text.

We now demonstrate the error correction process for the pair-complementary code. Consider the following codewords of $(\!(n,k)\!)$ pair-complementary code:
\begin{align}
    \ket{\bar{\psi}_u} &=\frac{1}{N_{u}} \left(
    (1-\gamma)^{-\frac{\|u\|}{2}}\ket{u00}+(1-\gamma)^{-(\frac{\|u\|+2}{2})}\ket{u11}
    -(1-\gamma)^{-\frac{n-\|u\|-1}{2}}\ket{\tilde u 10 }-(1-\gamma)^{-\frac{n-\|u\|-1}{2}} \ket{\tilde u 01} 
    \right)\; ; \nonumber \\
    \ket{{\bar\psi}'_u} &= \frac{1}{N'_{u}}\left(
    (1-\gamma)^{-\frac{n-\|u\|}{2}}\ket{\tilde u 11}+(1-\gamma)^{-\frac{n-\|u\|-2}{2}}\ket{\tilde u 00}
    +(1-\gamma)^{-\frac{\|u\|+1}{2}}\ket{u01}+(1-\gamma)^{-\frac{\|u\|+1}{2}}\ket{u10}
    \right) \;.
    \label{eqn:4-qubit cyclic codeword}
\end{align}
Here we use the superscript $u$ to label the pairs of logical states generated by the codeword $(\ket{u} + \ket{\bar u}/\sqrt 2)$ of the $(\!(n-2, k-1)\!)$ self-complementary code. And similarly as before, $N_{u}$ and $N'_{u}$ are normalization factors, and $\bar u$ is the complement of the bit-string $u$, where for each qubit site $i$, we have $\tilde u_i = 1- u_i$. 

For an input state $\ket{\bar\psi_\text{in}} = \sum_u a_u \ket{\bar{\psi}_u} + b_u\ket{\bar{\psi}'_u} $, there are three kinds of output states after $t=1$ amplitude damping, occurring with probabilities of order $O(\gamma)$ or higher. Firstly, an amplitude damping error with only $A^0$ acting on every single qubit changes the input state to,
\begin{equation}
    \ket{\phi_0}=\sum_u \frac{a_u}{N_{u}} \left(\ket{u00}+\ket{u11}-\ket{\tilde u 10}-\ket{\tilde u01}
    \right)+\frac{b_u}{N'_{u}} \left(\ket{\tilde u 11}+\ket{\tilde u 00}+\ket{u 01}+\ket{u 10 }
    \right)   \;.
\end{equation}

Secondly, for an amplitude damping error with operator $A^1$ acting on qubit site $i$, where $1 \leq i \leq n-2$, and operator $A^0$ acting on all other qubits, the input state transforms as:
\begin{equation}
    \ket{\phi_{1,i}} = \begin{cases}
        \sqrt{\frac{\gamma}{1-\gamma}}\sum_u \left[\frac{a_u}{N_{u}}(\ket{u'00}+\ket{u'11}) +\frac{b_u}{N'_{u}}(\ket{u'01}+\ket{u'10})\right]\;, \; \text{if}\ u_i=1 \;;\\
        \sqrt{\frac{\gamma}{1-\gamma}}\sum_u\left[\frac{-a_u}{N_{u}}(\ket{\tilde u'10}+\ket{\tilde u'01}) +\frac{b_u}{N'_{u}}(\ket{\tilde u'11}+\ket{\tilde u' 00})\right]
        \;, \; \text{if}\ u_i=0 \;.
    \end{cases}
\end{equation}
Note that this state contains half as many terms as $\ket{\phi_0}$. We denote by $u'$ the bit-string obtained by changing the bit at site $i$ from $1$ to $0$, and similarly define $\tilde{u}'$.

Finally, for an amplitude damping error with operator $A^1$ acting on one of the last two qubit sites and operator $A^0$ acting on all remaining qubits, the resulting output states are:
\begin{align}
  \ket{\phi_{\dots 10}} &= \sqrt{\frac{\gamma}{1-\gamma}}\sum_u \left[\frac{a_u}{N_{u}}\left(\ket{u01}-\ket{\tilde u 00}\right)  +\frac{b_u}{N'_{u}}\left(
  \ket{\tilde u01}+\ket{u00}
  \right)\right]\;, \\
  \ket{\phi_{\dots 01}} &= \sqrt{\frac{\gamma}{1-\gamma}}\sum_u \left[\frac{a_u}{N_{u}}\left(\ket{u10}-\ket{\tilde u 00}\right)  +\frac{b_u}{N'_{u}}\left(
  \ket{\tilde u10}+\ket{u00}
  \right)\right] \;.
\end{align}
Note that, similarly to the second case discussed above, these states contain half as many terms as $\ket{\phi_0}$. Furthermore, as discussed in Section~\ref{sec:4-qubit cyclic}, there is a large overlap between the error words $\ket{\phi_{\dots 10}}$ and $\ket{\phi_{\dots 01}}$. We therefore consider the classical mixture of these two error states and adopt the following set of projective measurements:
\begin{align}
    P_{0}=&\frac{1}{4}\left(
    \ket{u00}-\ket{\tilde u10}-\ket{\tilde u01}+\ket{u11} 
    \right) \left(
    \bra{u00}-\bra{\tilde u10}-\bra{\tilde u01}+\bra{u11} 
    \right) \nonumber\\
    &+\frac{1}{4}\left(\ket{\tilde u11}+\ket{u01}+\ket{u10}+\ket{\tilde u00} \right)\left(\bra{\tilde u11}+\bra{u01}+\bra{u10}+\bra{\tilde u00}
    \right) \; ,   \\
    P_{1,i}=&\begin{cases}
    \frac{1}{2}\left(\ket{u '00}+\ket{ u'11}
    \right)\left(\bra{u'00}+\bra{ u'11}
    \right)+\frac{1}{2}\left(
    \ket{ u'01}+\ket{ u'10}
    \right)\left(
    \bra{ u'01}+\bra{ u'10}
    \right)\;, \quad\text{if}\; u_i=0 \;;\\
    \frac{1}{2}\left(\ket{\tilde u '10}+\ket{\tilde u'01}
    \right)\left(\bra{\tilde u'10}+\bra{\tilde u'01}
    \right)+\frac{1}{2}\left(
    \ket{\tilde u'11}+\ket{\tilde u'00}
    \right)\left(
    \bra{\tilde u'11}+\bra{\tilde u'00}
    \right)\;, \quad\text{if}\; u_i=1\;,
    \end{cases}\;  \\
    P_S=&\frac{1}{6}\left(\ket{u01}+\ket{u10}-2\ket{\tilde u00}
    \right)\left(\bra{u01}+\bra{u10}-2\bra{\tilde u00}
    \right)
    \nonumber \\
    &+ \frac{1}{6}\left(\ket{\tilde u01}+\ket{\tilde u10}+2\ket{u00}
    \right)\left(\bra{\tilde u01}+\bra{\bar u10}+2\bra{u00}
    \right)\; ,
    \\
    P_A=&\frac{1}{2}\left(\ket{u01}-\ket{u10}
    \right)\left(\bra{u01}-\bra{u10}
    \right)+\frac{1}{2}\left(\ket{\tilde u01}-\ket{\tilde u10}
    \right)\left(\bra{\tilde u01}-\bra{\tilde u10}
    \right) \; .
\end{align}

We continue to calculate the worst-case fidelity achievable without residual logical errors.
To ensure clarity, we introduce the following notation. For simplicity of calculation, we denote the logical states by $\ket{\bar\psi_{\lambda}}$, labeled by index $\lambda$, which have the form:
\begin{equation}
\ket{\bar\psi_{\lambda}} = \frac{1}{N_\lambda}\sum_{\alpha}\pm(1-\gamma)^{-\|\lambda_\alpha\|/2}\ket{\lambda_\alpha}.
\end{equation}
Here, the index $\alpha$ labels the individual computational-basis components of the logical state, $N_\lambda$ is the normalization factor, and some bases carry alternating signs. If the noisy states are projectively measured to be in the subspace of $P_0$, after the error correction, the contribution to the fidelity is given by
\begin{equation}
    \mathcal{F}_0 = 4 \min_{\lambda}N_\lambda^{-2},
\end{equation}
which can be expanded in a Taylor series up to order $O(\gamma^2)$.
\begin{align}
     N_\lambda^{-2} &= \left[ \sum_\alpha(1-\gamma)^{-\|\lambda_\alpha\|} \right]^{-1}=\left[4+\left(\sum_\alpha\|\lambda_\alpha\| \right)\gamma+
    \frac{1}{2}\left(
    \sum_\alpha\|\lambda_\alpha\| +\sum_\alpha(\|\lambda_\alpha\|)^2
    \right)\gamma^2 +\cdots \right]^{-1} \nonumber \\
    &= \frac{1}{4}\left[
    1-\frac{\sum_\alpha \|\lambda_\alpha\|}{4}\gamma + \left(
    \frac{\left(\sum_\alpha \|\lambda_\alpha\|\right)^2}{16}-\frac{\sum_\alpha\left( \|\lambda_\alpha\|+ (\|\lambda_\alpha|)^2\right)}{8}
    \right)\gamma^2
    \right] +\cdots\;.
\end{align}
Note that from the structure of the pair-complementary codewords given in Eq.~(\ref{eqn:4-qubit cyclic codeword}), it follows that $\sum_{\alpha}\|\lambda_\alpha\| = 2n$ for any basis logical state.
Therefore, the leading contribution to worst-case fidelity is
\begin{equation}
    \mathcal{F}_0 = 1-\frac{n}{2}\gamma + \left(
    \frac{n^2}{4}-\frac{2n+\max_\lambda\left[\sum_\alpha  (\|\lambda_\alpha|)^2\right]}{8}
    \right)\gamma^2+O(\gamma^3) \;.
\end{equation}

Note that within each pair of the logical states in the pair-complementary codes, i.e. consider Eq.~(\ref{eqn:4-qubit cyclic codeword}) for a specific $u$, the sum of squared Hamming weights in each computational basis is the same, i.e. the quantity $\sum_\alpha (\|\lambda_\alpha\|)^2$ is equal. Thus to find the maximum, we only need to identify the bit-string $u$ that maximizes the following quantity $S(u)$, 
\begin{equation}
    S(u) =  \sum_\alpha (\|
\lambda_\alpha\|)^2 =\left[(\|u\|)^2+(\|u\|+2)^2+2(n-\|u\|-1)^2\right] \;.
\end{equation}
This maximum occurs when $u$ is the all-zero bit-string (or equivalently, the all-one bit-string), resulting in the maximum value:
\begin{equation}
   \max_u S(u)= 2^2+2(n-1)^2=  2n^2-4n+6 \;.
\end{equation}
Thus, we arrive at
\begin{equation}
    \mathcal{F}_0= 1-\frac{n}{2}\gamma +\frac{n-3}{4}\gamma^2 +O(\gamma^3)\;.
\end{equation}

We can similarly obtain the contribution from the projection $P_{1,i}$:
\begin{equation}
    \mathcal{F}_1 = \frac{2\gamma}{1-\gamma}  \min_\lambda N_\lambda^{-2} =\frac{\gamma}{2} 
    +(1-\frac{n}{2})\frac{\gamma^2}{2}+O(\gamma^3)\;.
\end{equation}

Finally, we can determine the contributions from the projections $P_S$ and $P_A$:
\begin{equation}
    \mathcal{F}_1' = \mathcal{F}_{1,S}'+\mathcal{F}_{1,A}' = (3+1) \cdot \frac{\gamma}{1-\gamma}\min_\lambda N_\lambda^{-2} =\gamma
    +(1-\frac{n}{2})\gamma^2+O(\gamma^3)\;.
\end{equation}

Summarizing these results, we obtain the worst-case fidelity up to the order $O(\gamma^3)$:
\begin{equation}
    \mathcal{F} = F_0 + (n-2)\mathcal{F}_1+\mathcal{F}_1'=1-\frac{n^2-3n+3}{4}\gamma^2+O(\gamma^3) \;,
\end{equation}
which is higher than the worst-case fidelity Eq.~(\ref{eq:f_nsa_sc_qubit}) of the $(\!(n,k)\!)$ self-complementary NSA code.

%%%%%%%%%%%%%%%%%%%%%%%%%%%%%%%%%%%%%%%%%%%
\section{Recovery and Fidelity of NSA Binomial Codes}

The binomial code has been proposed to use a single bosonic mode for boson loss and gain errors. In this section, we show that one can have an NSA generalization of the binomial code with improved fidelity. Since we focus on a single atom loss error in this paper, we consider the simplest version of the binomial code, i.e.\ the 0-2-4 code, which can correct the amplitude damping noise with $\ell\leq1$, where the Kraus operators are as follows:
\begin{equation}
    A^\ell=\left(\frac{\gamma}{1-\gamma}\right)^{\ell/2}\frac{a^l}{\sqrt{\ell!}}(1-\gamma)^{\hat{n}/2}\;.
\end{equation}
The code states are as follows:
\begin{align}
    |\bar 0\rangle&=\frac{1}{\sqrt{2}}(|0\rangle+|4\rangle)\; ;\nonumber\\
    |\bar 1\rangle&=|2\rangle\; .
\end{align}
Note that this can be thought of as being reduced from the $(\!(4, 1)\!)_2$
qubit code in~\cite{Leung_1997} by summing up the bit strings in each basis. Following our NSA generalization of the 4-qubit self-complementary code, we can generalize the 0-2-4 code to the NSA version, with the code states:
\begin{equation}
    |\bar{0}'\rangle=\sqrt{\frac{1}{1+(1-\gamma)^{-4}}}\left(|0\rangle+(1-\gamma)^{-2}|4\rangle\right)\;,\
    ~|\bar{1}'\rangle=|2\rangle\;.
\end{equation}

The recovery and fidelity of the (non)NSA 0-2-4 code is also similar to their (non)NSA 4-qubit counterpart, which we state below for concreteness. 

For the non-NSA 0-2-4 code, we consider the input qubit state, 
\begin{equation}
|\bar\psi_{\mathrm{in}}\rangle=a|\bar 0\rangle+b|\bar 1\rangle
\; ,
\end{equation}
after the amplitude damping noise $A^0$ or $A^1$ all possible final states will be:
\begin{align}
    |\phi_{0}\rangle&=a\frac{|0\rangle+(1-\gamma)^2|4\rangle}{\sqrt{2}}+b(1-\gamma)|2\rangle
    \; ;\\
    |\phi_{1}\rangle&=\sqrt{\frac{2\gamma}{1-\gamma}}(a(1-\gamma)^2|3\rangle+b(1-\gamma)|1\rangle)   \; .
\end{align}
The syndrome measurement could be done by a parity measurement of the boson number, and the fidelity after the recovery will be
\begin{equation}
\label{eq:fidelity_NonNSA_024}
    \mathcal{F}=\min\left\{\frac{1+(1-\gamma)^4}{2},(1-\gamma)^2\right\}+\frac{2\gamma}{1-\gamma}\min\{(1-\gamma)^4,(1-\gamma)^2\}=1-5\gamma^2+O(\gamma^3)
\end{equation}

For the NSA 0-2-4 code, we consider the input qubit state, 
\begin{equation}
|\bar \psi_{\mathrm{in}}\rangle=a|\bar 0'\rangle+b|\bar 1'\rangle
\; ,
\end{equation}
after the amplitude damping noise $A^0$ or $A^1$ all possible final states will be:
\begin{align}
    |\phi_{0}\rangle&=a\frac{|0\rangle+|4\rangle}{\sqrt{1+(1-\gamma)^{-4}}}+b(1-\gamma)|2\rangle
    \; ,\\
    |\phi_{1}\rangle&=\sqrt{\frac{2\gamma}{1-\gamma}}(\sqrt{\frac{2}{1+(1-\gamma)^{-4}}}a|3\rangle+b(1-\gamma)|1\rangle)   \; .
\end{align}
The syndrome measurement could also be done by a parity measurement of the boson number, and the fidelity after the recovery will be
\begin{equation}
    \mathcal{F}=\min\left\{\frac{2}{1+(1-\gamma)^{-4}},(1-\gamma)^2\right\}+\frac{2\gamma}{1-\gamma}\min\left\{\frac{2}{1+(1-\gamma)^{-4}},(1-\gamma)^2\right\}=1-3\gamma^2+O(\gamma^3)\;,
\end{equation}
which is higher than the non-NSA counterpart as in Eq.~(\ref{eq:fidelity_NonNSA_024}).

%%%%%%%%%%%%%%%%%%%%%%%%%%%%%%%%%%

%\bibliographystyle{apsrev4-2}
%\bibliography{reference.bib}
\bibliography{ref.bib}